\documentclass[fleqn,usenatbib,useAMS]{mnras}

\usepackage[pdftex]{graphicx}
\usepackage{epstopdf}
\usepackage{amsmath}	% Advanced maths commands
\usepackage{amssymb}	% Extra maths symbols
\usepackage{multicol}        % Multi-column entries in tables
\usepackage{bm}		% Bold maths symbols, including upright Greek
\usepackage{epsfig}
\usepackage{color}
\usepackage{multirow}
\usepackage{lscape}
\usepackage{pdflscape}
\usepackage{subfig}
\usepackage{placeins}

\newif\ifAMStwofonts

\def\degr{\hbox{$^\circ$}}
\def\arcmin{\hbox{$^\prime$}}
\def\arcsec{\hbox{$^{\prime\prime}$}}

\def\utw{\smash{\rlap{\lower5pt\hbox{$\sim$}}}}
\def\udtw{\smash{\rlap{\lower6pt\hbox{$\approx$}}}}

\title[V410 Puppis] {V410 Puppis: A useful laboratory for early stellar evolution}

\author[A. Erdem et al.]{Ahmet Erdem$^{1,2}$,   
Derya S\"{u}rgit$^{1,3}$,
Burcu \"{O}zkarde\c{s}$^{1,3}$,
Petr Hadrava$^{4}$,
Michael D. Rhodes$^{5}$,
\newauthor Tom Love$^{6}$,
Mark G. Blackford$^{6}$, 
Timothy S. Banks$^{7,8}$\thanks{Corresponding author: tim.banks@nielsen.com}, \& 
Edwin Budding$^{6,9,10,11}$
\vspace{2mm} \\
$^{1}${Astrophysics Research Center \& Ulupınar Observatory, \c{C}anakkale Onsekiz Mart University,} TR-17100, Çanakkale, Turkey\\
$^{2}${Dept.\ of Physics, Terzio\u{g}lu Kamp\"{u}s\"{u},} \c{C}anakkale Onsekiz Mart University, TR-17100, \c{C}anakkale, Turkey\\
$^{3}${Dept.\ of Space Sciences \& Technologies, Terzio\u{g}lu Kamp\"{u}s\"{u},} \c{C}anakkale Onsekiz Mart University, TR-17100, \c{C}anakkale, Turkey\\
$^{4}$Astronomical Institute of the Academy of Sciences of the Czech Republic, Boční II 1401/1, 141 00 Praha 4, Czech Republic\\
$^{5}${Department of Ancient Scripture, Joseph Smith Bldg, Brigham Young University, Provo, Utah 84602,} USA\\
$^{6}$Variable Stars South, {RASNZ, PO Box 3181, Wellington 6011, New Zealand} \\ 
$^{7}$Nielsen, 200 W Jackson Blvd, Chicago, {Illinois} 60606, USA\\
$^{8}${{Dept.\ of} Physical Science \& Engineering, Harper College, 1200 W Algonquin Rd, Palatine, {Illinois} 60067, USA}\\
$^{9}$Visiting astronomer, Mt.\ John Observatory,  Dept. of Physics \& Astronomy, Univ.\ of Canterbury, Private Bag 4800, Chch 8140, {NZ}\\
$^{10}$Carter Observatory, {40 Salamanca Road, Kelburn, Wellington 6012, New Zealand}\\
$^{11}$School of Chemical \& Physical Sciences, Victoria University of Wellington, {PO Box 600, Wellington 6140,} New Zealand\\
} 

\begin{document}
 
\pagerange{\pageref{firstpage}--\pageref{lastpage}} \pubyear{2022}
\maketitle
\label{firstpage}

\begin{abstract}
New spectrometric (HERCULES) and ground-based multi-colour photometric  data on  the multiple star V410 Puppis are combined with satellite photometry (HIPPARCOS and TESS), as well as historic astrometric observations. Absolute parameters for V410 Pup {Aab} are {derived}:  $M_{Aa}$ =  $3.15 \pm 0.10$, $M_{Ab}$ = $1.83 \pm 0.08$  (M$_{\odot}$); $R_{Aa}$ = $2.12 \pm 0.10$, $R_{Ab}$ =  $1.52 \pm 0.08$ (R$_\odot$); $a$ = $6.57 \pm 0.04$ R$_\odot$; $T_{Aa}$  = $12500 \pm 1000$, $T_{Ab}$ =  $9070 \pm 800$ (K), {and} photometric distance $350 \pm 10$ (pc). We report the discovery of a low-amplitude SPB variation in the  light curve and also indications of an accretion structure around  V410 Pup B as well as emission cores in V410 Pup C. We argue that V410 Pup is probably a young formation connected with the Vela 2 OB Association.  The combined evidence allows an age in the range 7-25 Myr from comparisons with standard  stellar evolution modelling.
\end{abstract}

%-------------------------------------------------------------------------

\begin{keywords} 
stars: binaries (including multiple) close --- stars: early type --- stars: individual V410 Pup 
\end{keywords}

%-------------------------------------------------------------------------

\section{Introduction}
\subsection{General}
\label{sec:introduction}  

The present article forms part of a `Southern Binaries Programme'  which in recent years has presented a number of studies of massive young close binary systems, often associated with star-forming regions. An interim summary was given by \cite{Idaczyk_2013}, and a recent example is the  paper of \cite{Budding_2021} on V Pup. Binary stars are an important source of fundamental data on stellar properties such as their masses and luminosities.  The ready availability of high-precision photometry from the TESS satellite (section~\ref{sec:photometry}) as well as access to the high-resolution HERCULES spectrometer (section~\ref{sec:spectroscopy}) allows ongoing refinement of our knowledge of such properties. The relatively fast development of massive stars  presents a special testing-ground for evolution mechanisms --- structural and dynamical. The importance of precision in the parametrization of stars has been stressed, for example in \cite{Eker_2015}.
  
{\cite{Popper_1980}'s classic paper  analysed 45 close binary examples to provide  detailed facts on stellar structure and evolution. He set parameter accuracy limits of  a few percent, which \cite{Andersen_1991} later improved to a cited $\sim 2\%$.} \cite{Andersen_1993} also drew attention to the interesting role close young binaries might play in understanding the relationship of stellar properties to their galactic environment (see also  \citealt{Southworth_2007}, \citealt{Gies_2013},  \citealt{Feiden_2015}).   \cite{Rucinski_2006} pointed out the observational neglect of binaries with declinations south of $-15^{\circ}$.  Such {discussions provide} a springboard for the work presented here. 

In the next subsection we review what is known about the V410 Pup system.  Section~\ref{sec:photometry} examines new and historic photometry, including data from the HIPPARCOS and TESS satellites as well as our new BVR light curves.  In section~\ref{sec:spectroscopy} we present observational work on data from the HERCULES spectrograph at the University of Canterbury Mt John Observatory in New Zealand.  In subsection~\ref{sec:velocity_determinations}, we  demonstrate radial velocity variations clearly identifiable with orbital motion of the close binary. Subsection~\ref{sec:mass_function} presents analysis of absorption line profiles to determine rotational angular velocities consistent with synchronized rotation and also raises the question of phase-dependence of the line equivalent width (the Struve-Sahade effect). Subsection~\ref{sec:korel} discusses application of the {\sc korel} disentangling technique and  to the complexities of this multiple star.  

The variable in  V410 Puppis is identified with the most massive A-component of a visual system containing at least three main objects: the B component having been discovered by R.\ Innes in 1906 and the C component by J.\ Herschel in 1836.  The separation of the AB-C system at $\sim$30 arcsec implies that there would be little, if any, detectable orbital motion in the years since discovery.  However, this is not the case for the AB binary, which  appears to have completed, since 1906, about one third of an eccentric orbit on the sky, whose  semi-major axis is $\sim$0.3 arcsec. 

Section~\ref{sec:astrometry} presents astrometric data and analysis on the AB system, allowing a check on derived masses.  Resulting absolute parameters are given in Section~\ref{sec:absolute_parameters} and in the final Section~\ref{sec:conclusion} these are compared with recent theoretical models published by the Padova school \citep{Bressan_2012}. While alternative scenarios for the age and evolutionary status of the system are possible, circumstantial  evidence favours a relatively young configuration. Further detailed observations are  recommended to improve our understanding of this interesting object.

\begin{table*}
\caption{Optimal parameters for the photometric model fits. {\sc WinFitter}  was used to model the HIPPARCOS, ground-based BVR, and TESS data from Sectors 9 and 35. Similar results were obtained for the Sector 7 data. Estimates from the Monte Carlo Wilson-Devinney program (WD + MC)  are given in the rightmost column, fitting the TESS data from Sector 35. These fittings specify near-equal `potentials' $\Omega_{Aa} = 3.743 \pm 0.033$, $\Omega_{Ab} = 3.736 \pm 0.077$. The mass ratios $q$ were fixed for each of these fittings. The limb-darkening requires an effective wavelength and this is determined from the assigned surface temperature $T_{\rm eff}$ values and  filter transmission data. {A mean wavelength of 800 nm  was taken} for the TESS photometry.  Coefficients ($u$) from the tables of \protect\cite{vanHamme_1993} are applied by {\sc WinFitter}. The quadratic formulation of \protect\cite{Claret_2017} was adopted in the  WD + MC technique, {with values being set to }$(X_{Aa}, Y_{Aa}) = (0.116, 0.238)$ and $(X_{Ab}, Y_{Ab}) = (0.186, 0.246)$. }
\label{tab:lc_fitting}
\begin{tabular}{lccccccc}
\hline  
\multicolumn{1}{c}{Parameter}  & \multicolumn{1}{c}{HIPPARCOS} & 
\multicolumn{1}{c}{B} & \multicolumn{1}{c}{V} & \multicolumn{1}{c}{R} &  \multicolumn{1}{c}{TESS Sec 9}  & \multicolumn{1}{c}{TESS Sec 35}   & \multicolumn{1}{c}{WD+MC}\\ 
\hline 
$M_{Ab}/M_{Aa}$            & 0.95               & 0.55                & 0.55               & 0.55               & 0.55              & 0.55                 & 0.58              \\     
$L_{Aa}$                   & 0.32 $\pm$ 0.06    & 0.54 $\pm$ 0.09     & 0.42 $\pm$ 0.10    & 0.29 $\pm$ 0.07    & 0.24 $\pm$ 0.04   & 0.305 $\pm$ 0.005    & 0.25 $\pm$ 0.02   \\  
$L_{Ab}$                    & 0.11 $\pm$ 0.04    & 0.20 $\pm$ 0.03     & 0.17 $\pm$ 0.08    & 0.17 $\pm$ 0.05    & 0.12 $\pm$ 0.03   & 0.110 $\pm$ 0.006    & 0.08 $\pm$ 0.01   \\  
$L_{B}$                     & 0.55 $\pm$ 0.06    & 0.26 $\pm$ 0.03     & 0.41 $\pm$ 0.05    & 0.53 $\pm$ 0.05    & 0.65 $\pm$ 0.04   & 0.574 $\pm$ 0.007    & 0.67 $\pm$ 0.02   \\ 
$r_{Aa} $ (mean)            & 0.297 $\pm$ 0.006  & 0.30 $\pm$ 0.01     & 0.30 $\pm$ 0.02    & 0.30 $\pm$ 0.02    & 0.295 $\pm$ 0.005 & 0.298 $\pm$ 0.001   & 0.32 $\pm$ 0.02   \\ 
$r_{Ab}$ (mean)             & 0.296 $\pm$ 0.01   & 0.29 $\pm$ 0.03     & 0.27 $\pm$ 0.05    & 0.29 $\pm$ 0.05    & 0.286 $\pm$ 0.013 & 0.276 $\pm$ 0.005    & 0.23 $\pm$ 0.01   \\ 
$i$ (deg)                               & 67.1  $\pm$ 1.0    & 66 $\pm$ 2          & 67 $\pm$ 3         & 70 $\pm$ 2         & 74.0 $\pm$ 0.5    & 70.3 $\pm$ 0.2       & 72.96 $\pm$ 0.36  \\ 
$T_{Aa}$ (K)                               & 11500              &      13000          &     13000          &      13000         &     13000         & 13000                & 12500             \\
$T_{Ab}$ (K)                               & 9000               &      9000           &      9000          &       9000         &      9000         & 9000                 & 9070              \\             
$u_{Aa}$                    & 0.48               & 0.44                & 0.38               & 0.31               & 0.48              & 0.48                 &                 \\   
$u_{Ab}$                    & 0.50               & 0.54                & 0.46               & 0.36               & 0.50              & 0.50                 &                  \\ 
$\chi^2/\nu$                            & 1.02               & 0.72                & 0.67               & 1.90               &  1.04             & 0.73                 & 1.13              \\ 
$\Delta l $                             & 0.007              & 0.01                & 0.01               & 0.01               &  0.001            & 0.001                 & 0.001              \\ 
\hline
\end{tabular}
\end{table*}

%{\bf There is agreement on the inclination and approximate sizes for Mark's LCs.  The 5relative luminosities are
%appreciably different, though there is general agreement that L1 becomes larger in B.
%The R data-set could be clipped of outliers a bit.  I think Mark deliberately 
%excluded 
%V410 C in his reductions, so we have seen L3 go down for these light curves. r2 has %come out a bit big in my fittings.  I wouldn't take that too seriously yet.
%}

%-------------------------------------------------------------------------------------------------

\subsection{V410 Pup}
\label{sec:background}

V410 Pup (HD 66079, HIP 39084, WDS J07598-4718AB)  is a magnitude 6.716 V (Tycho, see \citealt{Fabricius_2002})  multiple star, assigned a B8V spectral classification. It is located in the region of the Vela OB2 association, with a distance of $346 \pm 35$ pc (GAIA eDR3, \citeauthor{gaia_edr3}, \citeyear{gaia_edr3}).  The ICRS (2000) coordinates are 07h 59m 45.93 (RA) and $-47^{\circ}$ 18 \arcmin 12.7\arcsec (Dec), or galactic $\lambda$ 261.90 and $\beta$ 9.08 (deg).  {It was identified as} an eclipsing binary in the 74th name list of variable stars by \cite{Kazarovets_1999}, {based on} data from the HIPPARCOS survey \citep{de_Zeeuw_1999}. {\cite{Adelman_2000} used the same data to come to the same classification. This pair of stars is referred to as V410 Pup A, with the individual stars being Aa and Ab.}  The star is listed in the Catalogue of Eclipsing Binaries of \cite{Avvakumova_2013} as having a primary minimum depth of just 0.05 mag in V. V410 Pup AB-C, also known as HJ4032, was noted as a visual binary with a 9th magnitude companion at a separation of about 30 arcsec by John \cite{Herschel_1836}.  The 7.7 mag star {V410 Pup B was noted as part of the I 1070 \citep{Innes_1896} system, which we would now call V410 Pup AB. In summary, V410 Pup contains 4 known stars: 2 (Aa and Ab) as an eclipsing binary (A), and two other stars (B and C) in wide orbits with the original pair.}

{The system is reported in photometric and spectroscopic catalogs such as \cite{davis_1973}, \cite{Houk_1978}, \cite{Hauck_1978}, \cite{de_Bruijne_2012}, and \cite{Paunzen_2015}. While V410 Pup has been included in such catalogs, it has not before received a thorough study.}

Many stars in the Galaxy are thought to have originated in OB associations: large stellar groupings made conspicuous by their bright and massive young stars of early spectral type.  Detailed knowledge of the formation and properties of such `stellar nurseries' contributes to a broader understanding of the structure and history of the Galaxy.  Even so, available information remains largely summary in character, even for the nearest OB associations. This is partly because their large angular extent makes the identification of member stars challenging \citep{Preibisch_2006}.  Proving the membership of particular stars has sometimes been a protracted task. Considerable progress was achieved with the completion of the HIPPARCOS survey.   For reviews of the Vela OB2 Association see \cite{Jeffries_2009} and  \cite{Cantat-Gaudin_2019}.  Another recent study \citep{Kervella_2019} used information from Gaia to examine the incidence of binarity in the Vela OB2.

The {location} of a {complex multiple system such} as V410 Pup {relative} to Vela OB2 recalls that  of V831 Cen in relation to the nearer ($\sim 100$ pc), young ($\sim 10$ My)  Scorpius-Centaurus  OB2 association, where inclusion of astrometric analysis revealed additional parameters \citep{Budding_2010}.  
%Analysis of the light curves of V831 Cen supported %the presence of light contributors to the %photometry, additional to the two stars of the %eclipsing system. This gave rise to astrometric %analysis that revealed further parameters of the %V831 Cen system.  This parallels the situation we %find for V410 Pup.

%-------------------------------------------------------------------------

\section{Photometry}
\label{sec:photometry}
Since the discovery of variability of HIP 39084  was an outcome of the HIPPARCOS survey, examination of the relevant data in the HIPPARCOS Epoch Photometry Annex forms a natural starting place {for this project}.  \cite{esa_1997} reported the ephemeris {based on HIPPARCOS data as:} 
\begin{equation*}
\indent    {\rm Min} \,  {\rm I} = {\rm HJD} \, 2448500.3377 + 0.87617 E   \,  .
\end{equation*}
%Since the data were gathered over a $\sim$3 year interval after JD 2447862.0708 
% a more appropriate epoch for the HIPPARCOS coverage would be 
% HJD 2448455.5053; some 31 light cycles (45.090 d) earlier
% and close to the mean date of the data series.
%We adopted this for subsequent reference purposes.

\begin{figure}

\centering
\includegraphics[height=8.8cm]{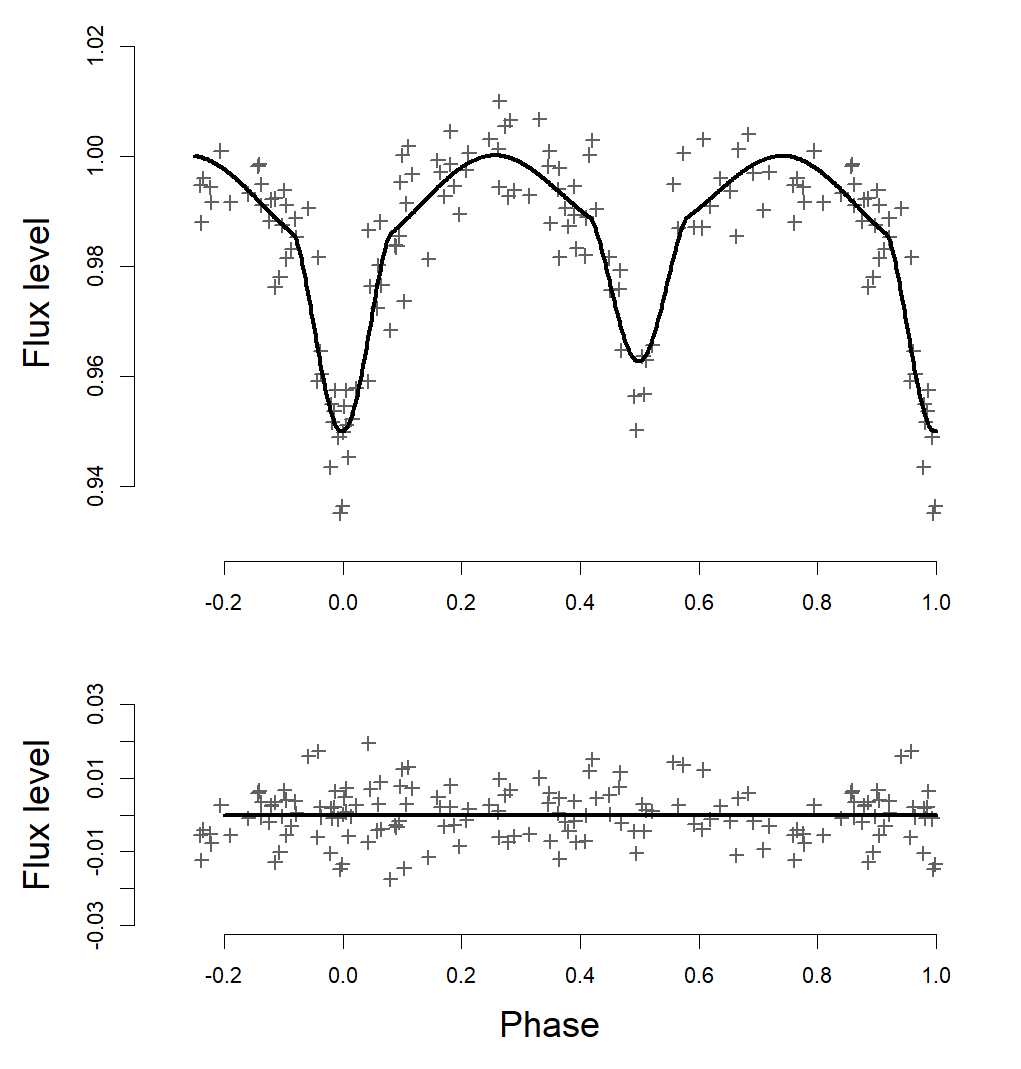}
\caption{HIPPARCOS photometry of V410 Pup with Radau-model fitting (shown as the solid line in the upper figure). \label{fig:hipparcos_lightcurve}
Residuals (crosses) are shown {in the lower figure, and distributed about zero on the y-axis.}}
\end{figure}

We applied the light-curve modelling program {\sc WinFitter}, which is described {in detail by} \cite{Rhodes_2021}. An early discussion of its optimization method was presented by \cite{Banks_1990}.  {\sc WinFitter} parametrizes a physically appropriate model that includes the effects of tidal distortion of the stellar envelopes (the {\em Radau} model, cf.\ \citeauthor{Kopal_1959}, \citeyear{Kopal_1959}). Its optimization procedures  are in keeping with the discussion of Bevington  (\citeyear{Bevington_1969}), particularly {his} chapter 11.   After locating an optimal parameter-set,  {\sc WinFitter}  numerically inverts the Hessian of the $\chi^2$ variate (in the vicinity of its minimum) {to derive the error matrix.} This Hessian should be positive definite for a properly posed data-modelling problem.  The resulting parameter uncertainties  then include the effects of inter-correlations between the {optimised parameters}. 

A preliminary {\sc WinFitter} model for the HIPPARCOS light curve is shown in Fig~\ref{fig:hipparcos_lightcurve} and a corresponding set of parameters listed in the first column of Table~\ref{tab:lc_fitting}. The appearance of the light curve together with the context of a young multiple star suggested  a pair of Main Sequence (MS) stars of closely comparable mass. The mass ratio was thus tentatively set at 0.95. {Stellar effective temperatures are required} to {be able to} parametrize the surface limb-darkening {values}. The listed B8 spectral type \citep{Wenger_2000} permitted  an initial estimate of $\sim$11500 K for the primary star Aa. Later determinations of the likely individual colours of the components caused an increase in this estimate for Aa to $\sim$13000 K. The ratio of eclipse depths is easily seen to equal the ratio of surface mean fluxes over corresponding eclipsed areas. The trend of the latter towards the effective temperature ratio in the Rayleigh-Jeans region of the spectrum applies to early type stars observed in the V and R ranges, allowing a rough estimate of $T_e \approx  9000$ K for Ab.

The Transiting Exoplanet Survey Satellite (TESS) \citep{Ricker_2014} has been operational since 2018.  V410 Pup was observed with 30-minute cadence by TESS in sectors 7 (January 7th to February 1st 2019), 8 (February 2 to February 27th 2019) and 9 (February 28th to March 26th 2019). More recently there have been observations with 10-minute cadence in sector 35 (February 9th to March 6th 2021).  We have processed these data using the {\sc Eleanor} Python package \citep{Eleanor_2019}. {These were first analysed using {\sc WinFitter}. We also applied the} numerical integration code of \cite{Wilson_1971} combined with the Monte-Carlo optimization procedure as discussed by \cite{Zola_2004}, {which is referred to} as {\sc WD + MC} in what follows. 

Although the form of the light curve was at first taken to be from a pair of  relatively similar unevolved and detached Main Sequence stars, the relative luminosities $L_{Aa, Ab}$, together with indications from the  spectroscopy (discussed in section~\ref{sec:spectroscopy}) pointed towards an early-type semi-detached arrangement, such as holds for V Pup (\citeauthor{Budding_2021}, \citeyear{Budding_2021}).

\begin{figure}
\centering
\includegraphics[height=6.5cm]{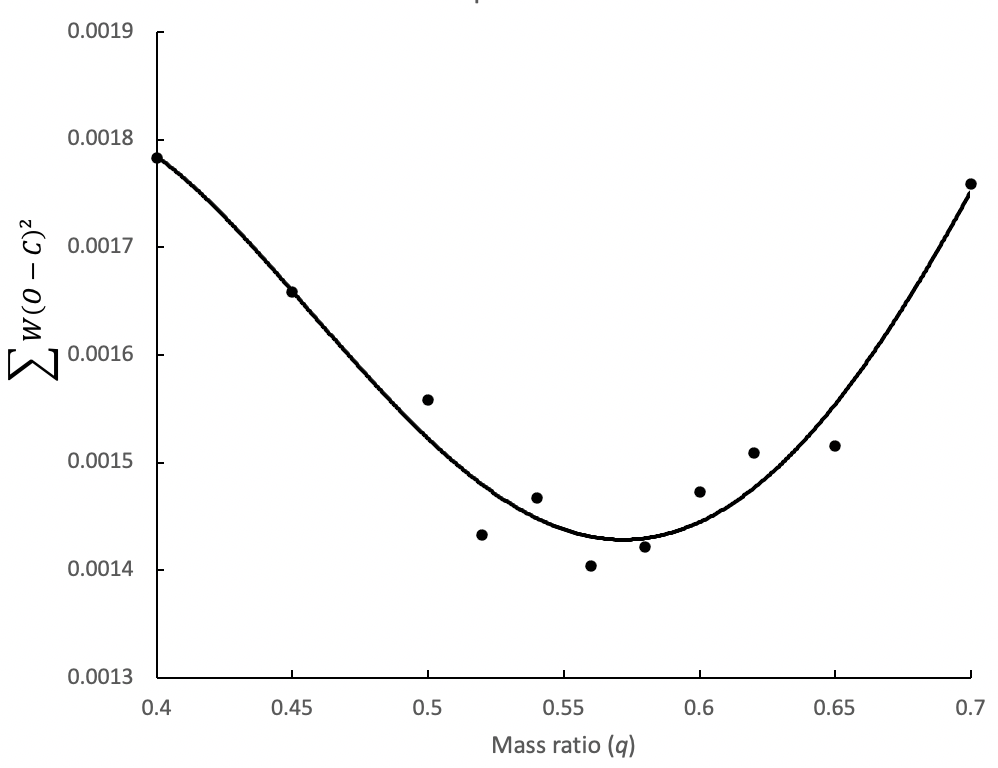} 
\caption{$\sum W(O-C)^2$ values corresponding to the optimal solutions by WD+MC for the range $0.4 < q < 0.7$, using Sector 35 TESS photometry of V410 Pup.  Similar searches were made using {\sc WinFitter} with the Sector 7 and 9 data, with comparable results.
\label{fig:sector_35_q_search} }
\end{figure}

\begin{figure}
\centering
\includegraphics[height=8.8cm]{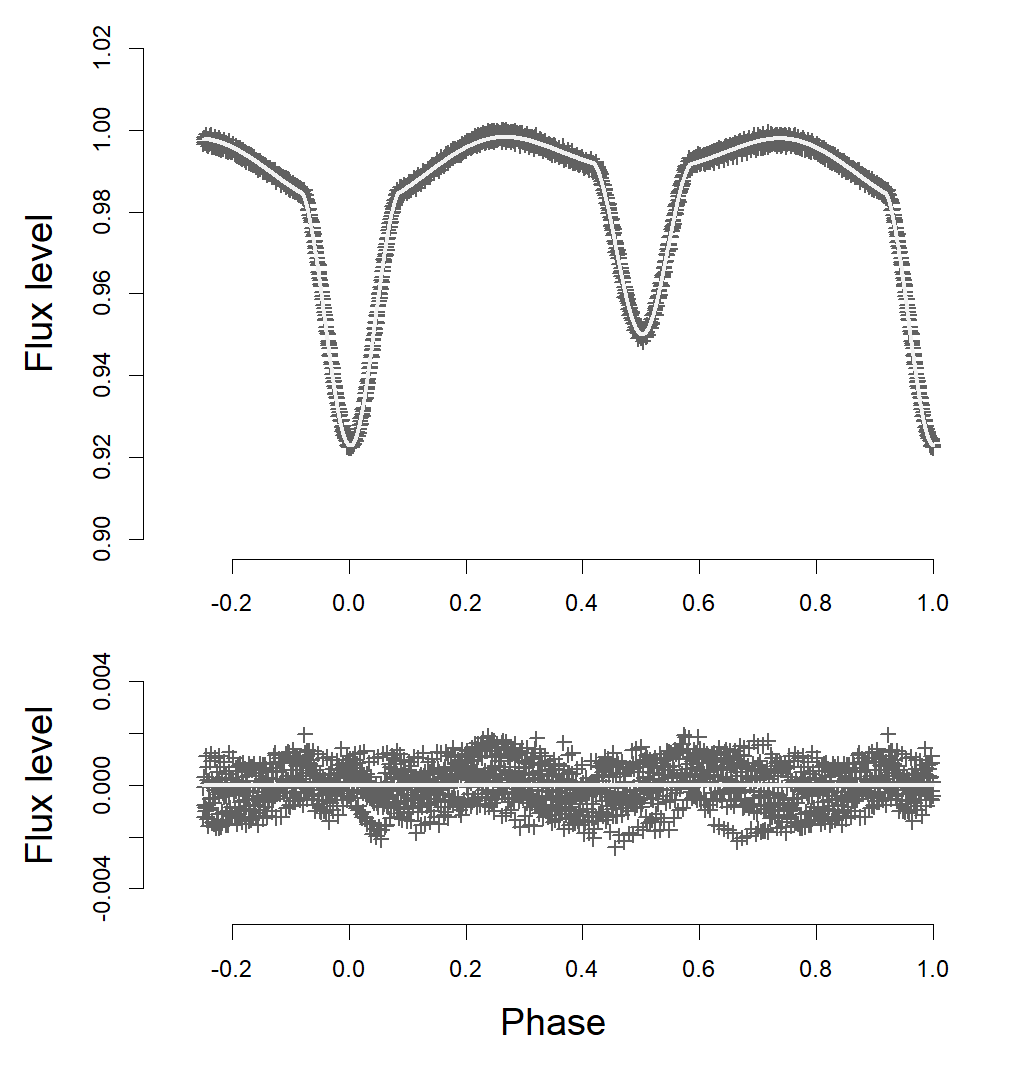}  
\caption{10-minute cadence (Sector 35) TESS photometry of V410 Pup with the optimal model from {\sc WinFitter} fitting. Residuals to the model are plotted in the lower figure.
\label{fig:TESS_lightcurve}}
\end{figure}

With this in mind we carried out a `q-search', allowing  optimal light curve fittings for given mass ratios ($q$) in the range  0.2 $< q <$ 1 to be checked.  A minimum $\chi^2$ was found at around $q \approx 0.58$ with the {\sc WD+MC} technique (Figure~\ref{fig:sector_35_q_search}), and similarly $q \approx 0.55$ for {\sc WinFitter}.
%The  parameter estimates from {\sc WinFitter} and {\sc WD + MC} fits with $q$ set %to 0.55 are given in Table~\ref{tab:lc_fitting}.  
The  parameter estimates from {\sc WinFitter} and {\sc WD + MC} fits with $q$ set to 0.55 and 0.58 are given in Table~\ref{tab:lc_fitting}. The best-fitting {\sc WinFitter} {model light curve for TESS Sector 35 data} is plotted in Fig~\ref{fig:TESS_lightcurve} against the {observations, together} with {a subplot showing} the residuals. 

\begin{figure}
\centering
\includegraphics[height=8.8cm]{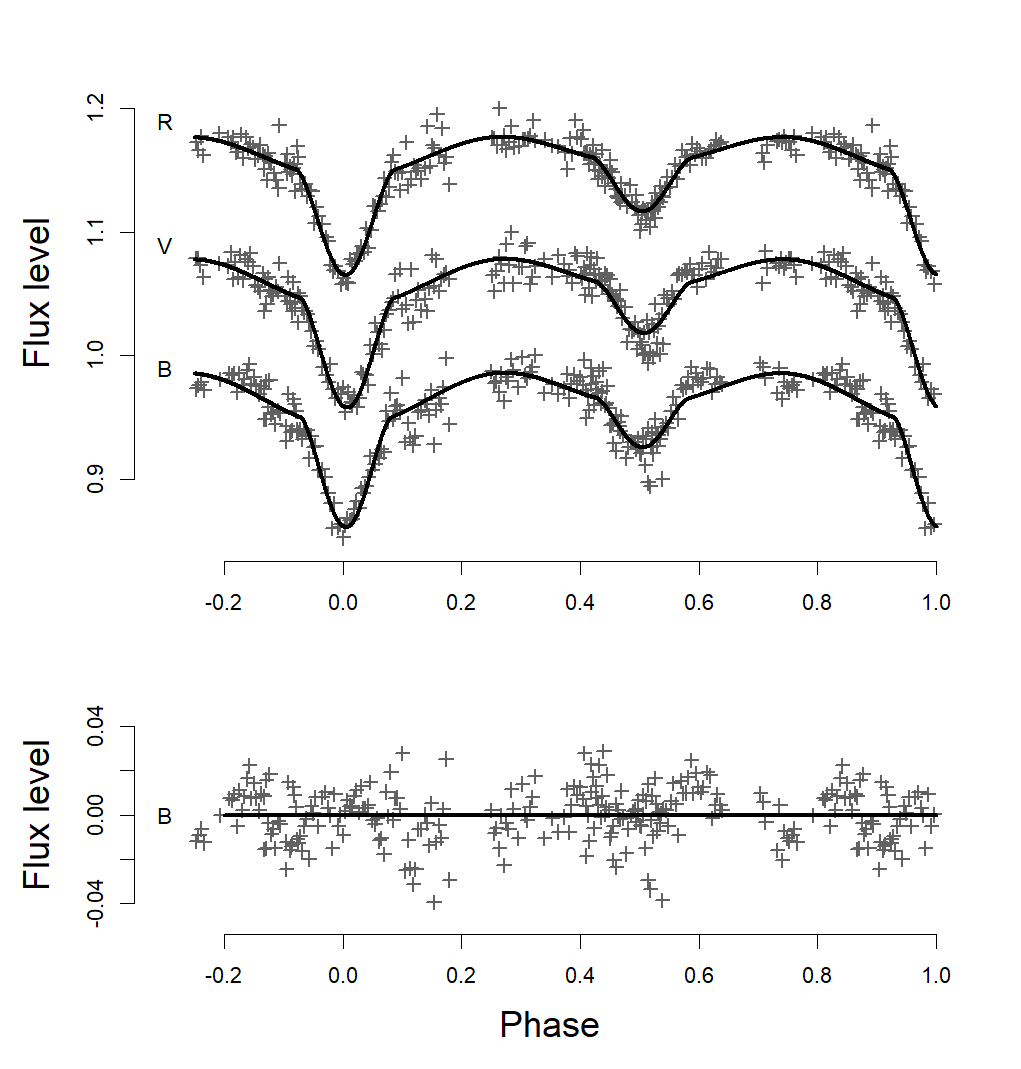}
\caption{BVR photometry of V410 Pup with Radau-model fitting.
Residuals (crosses) are shown distributed about the lower axis. The R data are offset vertically by $+0.2$, and the V band data by $+0.1$ in flux. Residuals are shown for  the B fit, being representative of the residuals for the other two bands.}
\label{fig:blackford_lightcurves}
\end{figure}

\begin{table}
    \caption{Magnitude and Colours of V410 Pup C and V410 Pup AB.}
    \centering
    \begin{tabular}{||c|r|r|r|r||}
    \hline
          & \multicolumn{2}{c}{V410 Pup AB} & \multicolumn{2}{c}{V410 Pup C} \\
          & Magnitude & Error & Magnitude   & Error \\
    \hline
    $V$   & 6.716     & 0.024 & 9.461       & 0.021 \\  
    $B-V$ & $-0.056$  & 0.034 & 0.074       & 0.030 \\
    $V-R$ & $-0.014$  & 0.035 & 0.028       & 0.031 \\ 
    \hline
    \end{tabular}
    \label{tab:mark_c_separate}
\end{table}

\begin{table}
    \caption{Eclipse timings based on the BVR photometry.  HJD is Heliocentric Julian Date.}
    \centering
    \begin{tabular}{||c|c|l||}
        \hline
         HJD         & Error (days) & Eclipse type \\
        \hline
        2458488.0254 & 0.0061       &  Secondary \\
        2458494.1599 & 0.0051       &  Secondary \\
        2458548.0423 & 0.0039       &  Primary \\
         \hline
    \end{tabular}
    \label{tab:mark_eclipse_timings}
\end{table}

New photometric observations of V410 Pup  were obtained over three nights in January 2019 using an 80mm F6 refractor telescope, stopped down to 50-mm aperture, at the Congarinni Observatory, NSW, Australia ($152\degr$ $52\arcmin$ E, $30\degr$ $44\arcmin$ S, 20 meters above mean sea level). The light curves are shown in Figure~\ref{fig:blackford_lightcurves}. The images were captured by an ATIK$^{\rm TM}$  One 6.0 {Charge-Coupled Detector (CCD)} camera equipped with Johnson-Cousins {\em BVR} filters. Further photometric observations were made on one night in March 2019 using the same equipment with the addition of a $2\times$ tele-extender to increase the image scale, allowing measurement of V410 Pup C separately from V410 Pup AB (see Table~\ref{tab:mark_c_separate}). The comparison star was HD65817 ($V = 8.223$, $B-V = -0.021$, $V-R = 0.004$). The {\sc MaxIm DL}$^{\rm TM}$ software package \citep{George_2000} was used for image capture, calibration and aperture photometry. 

We determined {1 primary and} 2 secondary times of minimum (see Table~\ref {tab:mark_eclipse_timings}) {via polynomial fitting available in} the {\sc peranso$^{\rm TM}$ } package \citep{Paunzen_2016}.  The results of applying {\sc WinFitter} to this standard {{\em BVR}} photometry are shown in Fig~\ref{fig:blackford_lightcurves}, and key parameters tabulated in Table~\ref{tab:lc_fitting} together with the earlier fitting results.  The relatively long time baseline HJD 2448500.3377 to 2458548.0423 permitted an improved estimate of the mean period as $P =  0.8761514$ d.  This shorter period reduced some of the apparent scatter in the data phasing.  

A key result coming from the new photometry and the modelling results presented in Table~\ref{tab:lc_fitting} is that the proportion of background light increases significantly with increasing wavelength.  {Fainter optical sources} within the ABC system's separation are visible in the SDSS imaging \citep{Blanton_2017}.  These may be cooler stars adding light into the {21-arcsec} wide pixels of the TESS {CCDs}.  To pursue this, Gaia's eDR3 database was checked, covering the sky around V410 Pup within a 1 arcminute radius.  Although over 200 sources were thus located, they are, by far, stars with V mag fainter than 15. The only listed stars brighter than 13th magnitude are the AB and C components of V410 Pup. We thus rule out field stars, and look only within the multiple star itself for the source of increasing brightness at longer wavelength. Given the relative faintness of star C (Table \ref{tab:mark_c_separate}), this extra light can be reasonably associated with star B (cf.\ Table~\ref{tbl-4}).

\begin{table}
    \caption{$ B V R$ Magnitudes of V410 Pup Aa, Ab and B.}
    \centering
    \begin{tabular}{|c|c|c|c|c|}
    \hline
 \multicolumn{1}{c}{Source}
          & B  & V   & R & Error \\
    \hline
    $A_a$  & 7.48  & 7.66 & 7.83 & 0.02       \\  
    $A_b$  & 8.35  & 8.41 & 8.40 & 0.04      \\
    $B$    & 8.00  & 7.80  & 7.65 & 0.03       \\ 
    \hline
    \end{tabular}
    \label{tbl-4}
\end{table}

Component Aa is clearly a blue star, attaining a typical mid-B type colour when interstellar reddening is taken into account.  Using the procedure of \cite{Cardelli_1989}, with the adopted distance (see Section~\ref{sec:background}), implies we subtract 0.09 from the B -- V value of $-0.18$ deduced from Table~\ref{tbl-4}.

Component Ab appears to have the zero colour appropriate to an A0 type spectrum, although its relative luminosity in comparison to Aa is high for both to be typical Main Sequence stars. Component B looks to have colours corresponding to a temperature of $\sim$7500 K, but it is also too bright to be a Main Sequence star. {It also displays} increasing luminosity towards longer wavelengths.

\begin{figure}
\centering
\includegraphics[height=6cm]{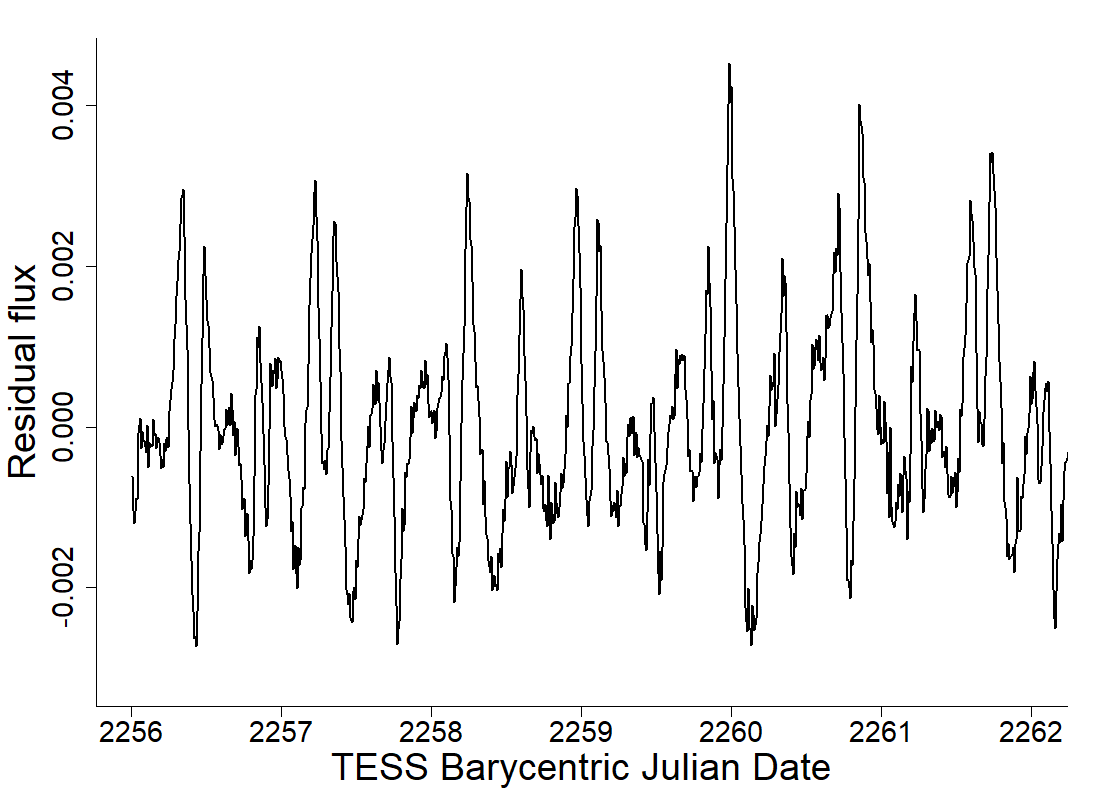}
\caption{Pulsations observed in the residual light curve of V410 Puppis once the modelled eclipsing binary light curve has been removed.}
\label{fig:pulse} 
\end{figure}

\begin{figure}
\centering
\includegraphics[height=6cm]{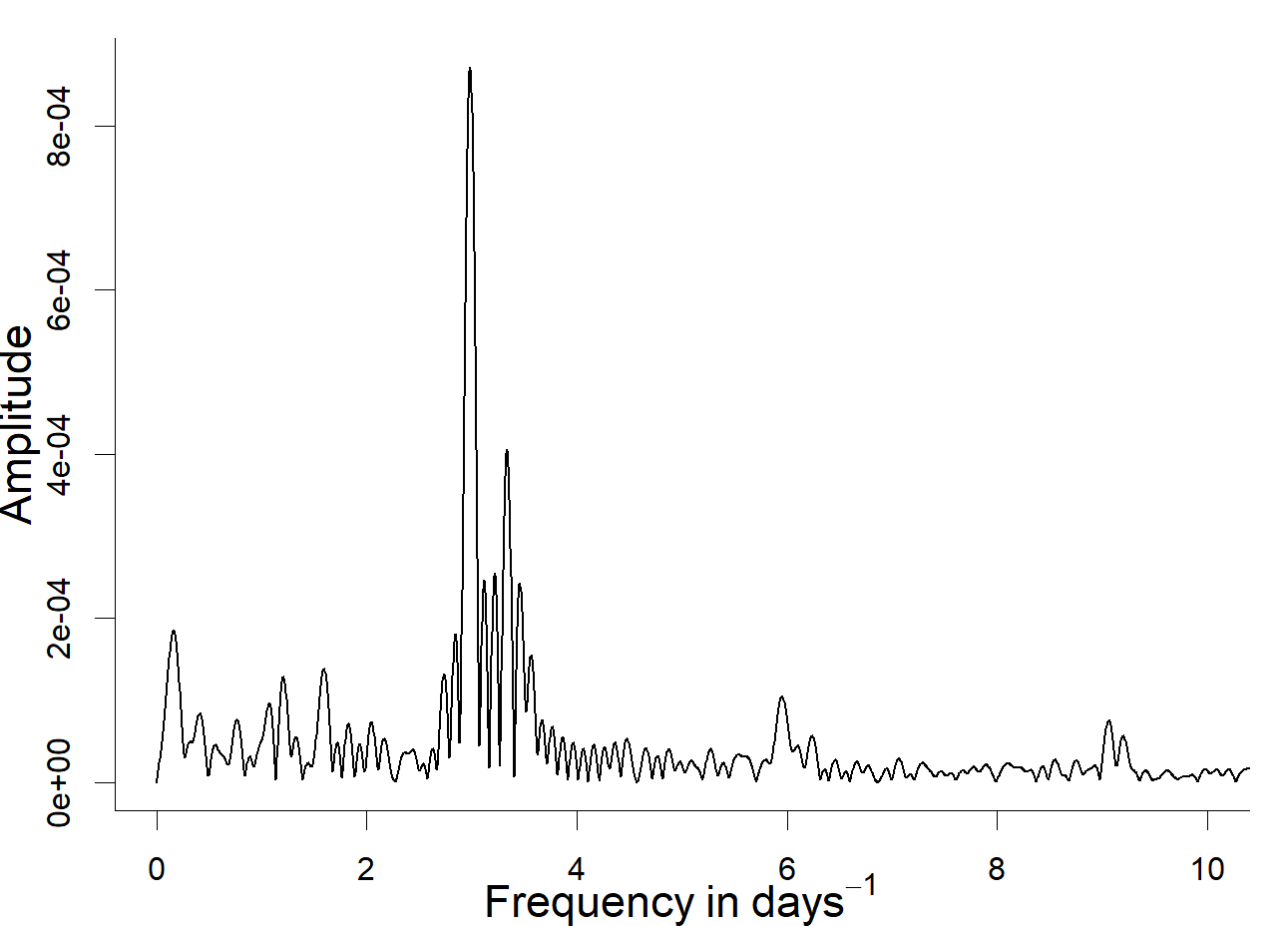}
\caption{Frequency plot of pulsations after removal of harmonics of the orbital frequency.}
\label{fig:fourier} 
\end{figure}

Figure~\ref{fig:TESS_lightcurve} {(on page~\pageref{fig:TESS_lightcurve})} shows a thick band of residuals in the phased TESS high cadence lightcurve data, with a greater amplitude than should be expected from the precision of TESS data.  After fitting the eclipsing model we converted the phased residuals of the lightcurve back to an absolute time series.  The resulting residual plot appeared to show regular variation over timeframes of a few hours (Figure~\ref{fig:pulse}). We used {\sc period04} software \citep{Lenz_Breger_2004} to pre-whiten the frequency plot for harmonics of the orbital period, leaving a significant frequency detection at  2.98/d, or $\rm P = 8.05$ hours (Figure~\ref{fig:fourier}).  This period could be consistent with the existence of a Slowly Pulsating B star (SPB) in the V410 Pup system, potentially similar to the SPB HD 111774 in the Sco-Cen association  \citep{Sharma_2022}.  All components of the V410 Puppis system lie within the aperture used for TESS photometry, and it is therefore not clear from that photometry which member of the system is the pulsating star, or whether more than one star in the system might be exhibiting pulsations.  However, we believe that the Aa star is most likely to exhibit SPB phenomena, given the likely evolutionary status of the stars discussed in Section~\ref{sec:conclusion} below.

%---------------------------------------------------------------------------------------------

\subsection{Orbital Period}
\label{sec:orbital_period}

The possibility of a semi-detached configuration for V410 Pup  {motivated} checking {the} times of minimum light (ToMs), {leading to} a comparison of {these} observed ToMs with those calculated from an adopted linear ephemeris ({i.e., observed minus calculated, abbreviated as} $\rm O - C$s). Although the system is relatively bright, no ToMs {have been} published {in the literature}. Relevant photometric data are found in the HIPPARCOS, ASAS-3 (see \citealt{Pojmanski_2002} for background), OMC \citep{Alfonso_Garzon_2012}, and TESS data-bases.  ToMs can be {calculated directly} from the TESS photometry, but observations in the other data-bases usually contain only one or two points on a given night. We therefore employed a method similar to that of \cite{Zasche_2014} to derive appropriate ToMs.  {This led to} four times of minima from HIPPARCOS, 7 from ASAS-3, 3 from OMC, {and} 5 from TESS {being obtained in addition to the three reported above.}  Many more individual ToMs could have been taken from the full range of TESS data, but it is sufficient here to determine only representative minima {given the short duration of a TESS sector.}

\begin{figure}
\centering
\includegraphics[height=11cm]{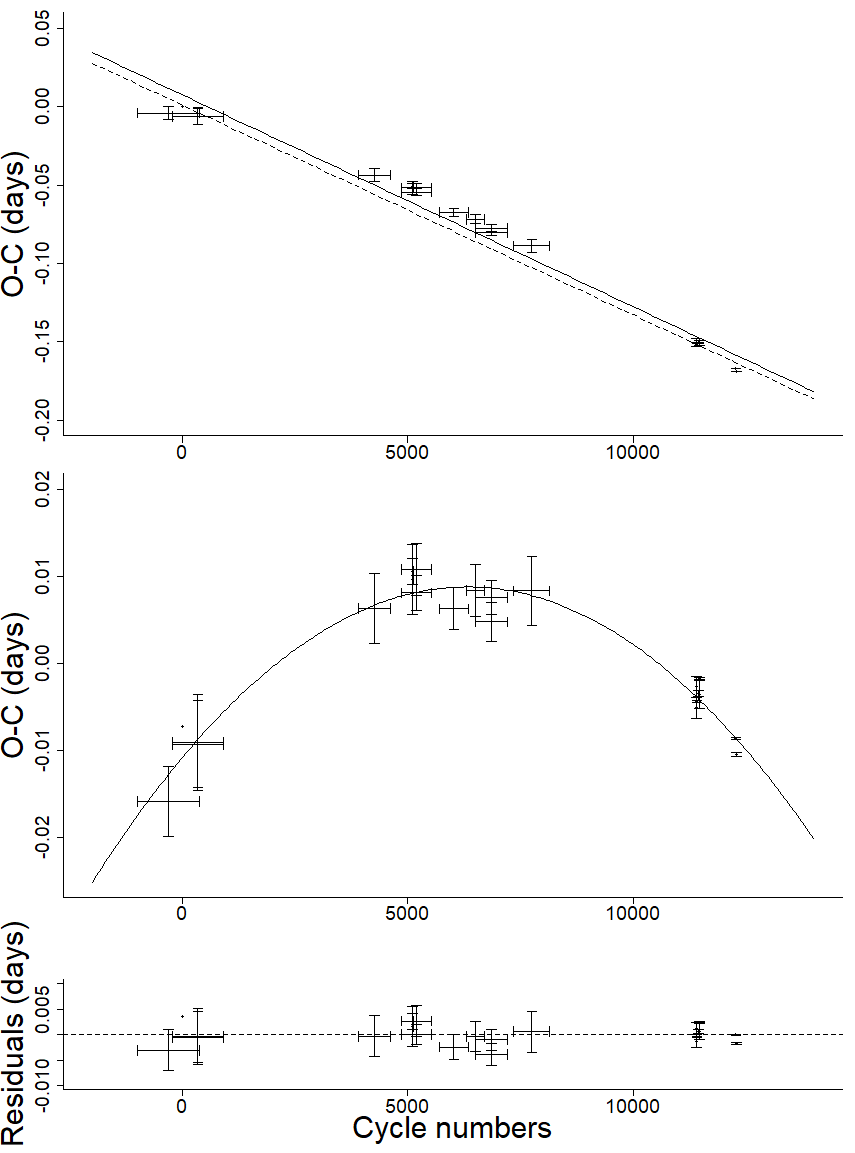}
\caption{O--C trend using historical data and the TESS 35 ToMs discussed in section~\ref{sec:orbital_period}.  Top panel: Linear fit.  Middle panel: quadratic fit.  Bottom panel: quadratic residuals}
\label{fig:ObsCalc} 
\end{figure}

Using the light elements from HIPPARCOS and these 22 ToMs,  ($\rm O - C$)s were calculated and plotted against the {(orbital)} cycle number (cf.\ the top diagram of Figure~\ref{fig:ObsCalc}). The diagram shows the result from optimizing a linear model to derive a representative mean ephemeris. The linear model for all the ($\rm O - C$)s is shown with a solid line, which can be compared with the best-fit linear model for only the TESS ToMs (dashed line). The {following equation was derived}:
\begin{equation}
T_{\rm min} \, {\rm (HJD)}  = 2448500.3377(30) + 0.8761516(3) \times E
\label{eq:one}
\end{equation}
The period here is noticeably shorter than the HIPPARCOS value given at the beginning of this section.

From the top diagram in Fig~\ref{fig:ObsCalc} it is seen that the ($\rm O - C$) behaviour has  some additional systematic variation. A second-order polynomial model  was then least-squares fitted to this variation. The corresponding model is shown in the middle diagram of Figure~\ref{fig:ObsCalc}, {corresponding to the} equation:
\begin{eqnarray}
T_{\rm min} \, {\rm (HJD)} 
& = & 2448500.3334(10) + \big ( 0.8761577(3) \times E \big ) \nonumber \\
 &  & – \: \big ( 4.90(26) \times 10^{-10 } \times E^2 \big ) \nonumber 
\end{eqnarray}
\noindent
The quadratic term reveals that the orbital period of V410 Pup is decreasing at the relatively large rate of 0.035 $\pm$ 0.002 s/year  over the last $\sim$ 30 years, implying significant loss of orbital  angular momentum. This may result from various mechanisms. {A close binary context, such as for this system, suggests} interactive evolution.  Alternatively, given the primary's early spectral type, hot stellar winds may remove angular momentum from the system.  A classical Algol configuration, hinted at in the beginning of this subsection, would be more often expected to show period lengthening, i.e., an upturned parabola in the $\rm O - C$  diagram, opposite to that of Fig~\ref{fig:ObsCalc}, though  that is not always the case (cf.\ Chapter 8 of \citealt{budding_demircan}).

%-------------------------------------------------------------------------

\section{Spectroscopy}
\label{sec:spectroscopy}

The High Efficiency and Resolution Canterbury University Large \'{E}chelle Spectrograph (HERCULES, \citealp{Hearnshaw_2002}) provided our spectrosopic data.  HERCULES {was used} with the 1m McLellan telescope at the University of Canterbury Mt.\ John Observatory (UCMJO) near Lake Tekapo ($\sim$ 43{\degr}59{\arcmin} S, 174{\degr}27{\arcmin} E).  Images were collected with a 4k$\times$4k Spectral Instruments (SITe) camera. 
%\cite{Skuljan_2014} {\bf There may be a more appropriate reference  here -- if necessary.}
Starlight was passed from the telescope to the spectrograph through a 100 $\mu$ fibre equipped  with a  microlens at the fibre entrance, that produces an entrance pupil of effectively 4.5 arcsec diameter.  This is suited to typical seeing conditions at Mt.\ John. An instrumental resolution of $\sim$40000 is generally attained. The normal procedure for wavelength and relative flux calibration was followed (cf. \citeauthor{Blackford_2019}, \citeyear{Blackford_2019}). 
% however,
%an error in the detector's flux meter led to under-exposure 
%of some calibration frames.  Subsequent checks on telluric lines in the science 
%frames provided assurance that our final data reductions were accurate and consistent.
Exposure times were usually about 500 seconds in mostly clear weather. 

%-------------------------------------------------------------------------------------------

\subsection {Spectrum measurements}
\label{sec:velocity_determinations}

33 spectra  of V410 Pup were selected from spectral images taken over the years 2011-15 and early 2020.  A recent version of the software package {\sc hrsp} \citep{Skuljan_2020} was used to produce wavelength calibrated and normalized output in {\sc fits} formatted files. About 40 useful spectral orders in the image plane were examined, covering the region of  450 to 700 nm. The spectra are mostly without strong features, apart from highly broadened lines of H$_{\alpha}$ and H$_{\beta}$, with the former affected by telluric intrusions. Spectra were inspected using {\sc vspec} \citep{Desnoux_2011}. Broad and shallow He I lines are typically observed for this type of hot close binary system. {In the present instance the additional light of V410 Pup B,} as well as the late B classification \citep{Houk_1978}, {rendered} the primary poorly measurable {and} we were not able to identify secondary helium lines with any confidence. Tentatively identified features are listed in Table~\ref{tab:spectral_features}.  The He I lines had sufficient definition to enable velocity estimation through the Doppler-shift principle, though the blend at $\lambda$ 4922~\AA\ introduces complications, causing us to neglect this feature in these direct radial velocity (RV) determinations.

\begin{table}
\begin{center}
\caption{Identified spectral lines for V410 Pup 
{based on comparison with synthetic spectra from \protect\cite{Gummersbach_1996}.} 
Only {primary (Aa)} lines are confidently detected except for the H lines, whose varying widths and asymmetries evidence the secondary (Ab) contribution.
\label{tab:spectral_features}}
%{\footnotesize
\begin{tabular}{lcll}
%  & & &  \\
  \hline 
\multicolumn{1}{l}{Species}  & \multicolumn{1}{c}{Order no.} &
\multicolumn{1}{l}{Adopted $\lambda$} & \multicolumn{1}{l}{Comment}  \\
\hline
%Fe I      *   & 81        &  7024.7  &   (ident.\ ?)     \\
%(sol)        & 82        &  6924.4  &        \\
He I         & 85        &  6678.1    &   measurable Aa   \\
H$_{\alpha}$ & 87        &  6562.8    &   strong, blended Aa \& Ab       \\
Fe II        & 89        &  6376.0     &  very weak  \\
{Si} II        & 90        &  6347.8      &  broad \& shallow blend \\  % may be blended with TiI features
%N II      *   & 96        &  5940.3  &   \,\,\, ''  \,\,\, ''   \\
He I         & 97       &  5875.7     &  measurable Aa\\
%C III        & 104      &  5457.2     & should be there, but very weak\\
Fe II         & 110      &  5169.0     &  very weak   \\
Si II         & 112      &  5056.0     & weak blend  \\
He I/Si II    & 115      &  4921.9    & detectable  Aa   \\
H$_{\beta}$   & 117      &  4861.3    &  strong, blended Aa \& Ab      \\
Fe II         & 124      &  4583.8     &  very weak     \\
Fe II         & 125      &  4549.5     &  weak blend   \\
%Si III        & 125      &  4552.65  & blend     \\
\hline \\
\end{tabular}
\end{center}
\end{table}

%The phase coverage of these UCMJO observations was about 65\% of the complete cycle.
%Uncovered phases have been observed subsequently (TL) from the Saesteorra Observatory in the Wairarapa %region of New Zealand 
%(41$^{\circ}$ 13$^{\prime}$ 07$^{{\prime}{\prime}}$S;  
%175$^{\circ}$ 27$^{\prime}$ 51$^{{\prime}{\prime}}$
%E) with a 12 inch f/8 Ritchey-Chretien telescope.  The spectroscopic equipment there
%consisted of a Lhires Littrow design spectrograph (Thizy, 2007), 
%having a 35 micron slit and a 2400 lines/mm grating.  This has a nominal resolution of 
%approximately R=13400. 
%{\bf The spectral coverage was set to be from 6607 to 6813 {\AA}, 
%so as to encompass the He6678 line. }

\begin{table}
\begin{center}
\caption{Radial velocity data of the primary component from the He I lines. Phase was calculated using equation~\ref{eq:one}.
\label{tbl-7}} 
\begin{footnotesize}
\begin{tabular}{llrll} 
\hline

\multicolumn{1}{l}{BJD}  & \multicolumn{1}{l}{Orbital} & 
\multicolumn{1}{r}{RV1} & \multicolumn{1}{l}{$\sigma$} & \multicolumn{1}{l}{EW}  \\
\multicolumn{1}{l}{2450000+}  & \multicolumn{1}{l}{phase} &
\multicolumn{1}{r}{km s$^{-1}$} & \multicolumn{1}{l}{km s$^{-1}$}\\ 
%&  \multicolumn{r}{r}{ }\\
\hline 
  5879.9288	& 0.733	&	    165	   & 15         & 0.131  \\
  5880.1346	& 0.968	&	 	 83	   & 10         & 0.120  \\
  5880.8750	& 0.813 &		163	   & 20         & 0.196  \\
  5880.9035	& 0.846 &		155    & 20         & 0.131  \\
  5883.0634	& 0.311 &	   $-73$   & 10         & 0.127  \\
  5883.0845	& 0.335 &	   $-65$   & 8          & 0.127  \\
  5883.1259	& 0.382 &	   $-40$   & 8          & 0.118  \\
  6257.8766	& 0.106 &	   $-36$   & 8          & 0.161  \\
  6257.9458	& 0.185 &	   $-78$   & 10         & 0.131  \\
  6259.0672	& 0.465 &	   $-10$   & 5          & 0.121  \\
  6259.1049	& 0.508 &		 24	   & 5          & 0.123  \\
  6667.0650	& 0.135 &	   $-57$   & 8          & 0.117  \\
  6669.9055	& 0.377 &	   $-51$   & 8          & 0.153  \\
  6669.9385	& 0.415 &	   $-36$   & 6          & 0.128  \\
  6670.8979	& 0.510 &		 25	   & 5          & 0.021  \\
  6671.8944	& 0.647 &	    110	   & 15         & 0.087  \\
  6673.9285	& 0.969 &		 76	   & 8          & 0.077  \\
  6994.1309	& 0.434 &	   $-24$   & 5          & 0.098  \\
  6998.9403	& 0.927 &	    110	   & 10         & 0.074   \\
  7006.0139	& 0.996 &		 65	   & 18         & 0.022   \\
  7351.1053	& 0.869 &	    135	   & 20         & 0.077   \\
  7352.0590	& 0.957 &	     79	   & 7          & 0.078  \\
  7355.0698	& 0.393 &	   $-42$   & 5          & 0.088  \\
  7356.0614	& 0.525 &		 41	   & 6          & 0.071  \\
  7357.1201	& 0.737 &		153	   & 15         & 0.096  \\
  7358.0074	& 0.746 &	    155	   & 15         & 0.093  \\
  7359.9993 & 0.020 &	     21	   & 5          & 0.108  \\
  7360.9513 & 0.106 &	   $-25$   & 10         & 0.141  \\
  8887.9102 & 0.909 &		 87	   & 15         & 0.096  \\
  8887.9225 & 0.927 &		 74    & 6          & 0.093  \\
  8887.9729	& 0.981 &		 62	   & 8          & 0.091  \\
  8887.9831	& 0.992 &	 	 42    & 8          & 0.073 \\
  8890.9390 (C)	& --- &	     29    & 2          & ---   \\
\hline 
\end{tabular}
\end{footnotesize}
\end{center}
\end{table} 

%-------------------------------------------------------------------------------

%\subsection{Rotational velocities}

If the resolution is sufficiently high, spectral line profiles can be modelled with a parameter set that determines the radial velocity of the centre of light, as well as the  source's bodily rotation and the scale of turbulence in the surrounding plasma. Such modeling has been carried out at least since the work  of \cite{Shajn_1929}; for a review, see \cite{Collins_2004}. \citeauthor{Shajn_1929}'s model involved the convolution of  rotational and Gaussian broadening functions.  Such a fitting function was used within the optimization program {\sc prof}, introduced in \cite{Olah_1992} (see also \citeauthor{Budding_1994}, \citeyear{Budding_1994}). Application of {\sc prof} to the He I $\lambda$ 6678 and $\lambda$ 5876  features resulted in the RV values listed in  Table~\ref{tbl-7}.  

\begin{figure}
\centering
\includegraphics[height=7cm]{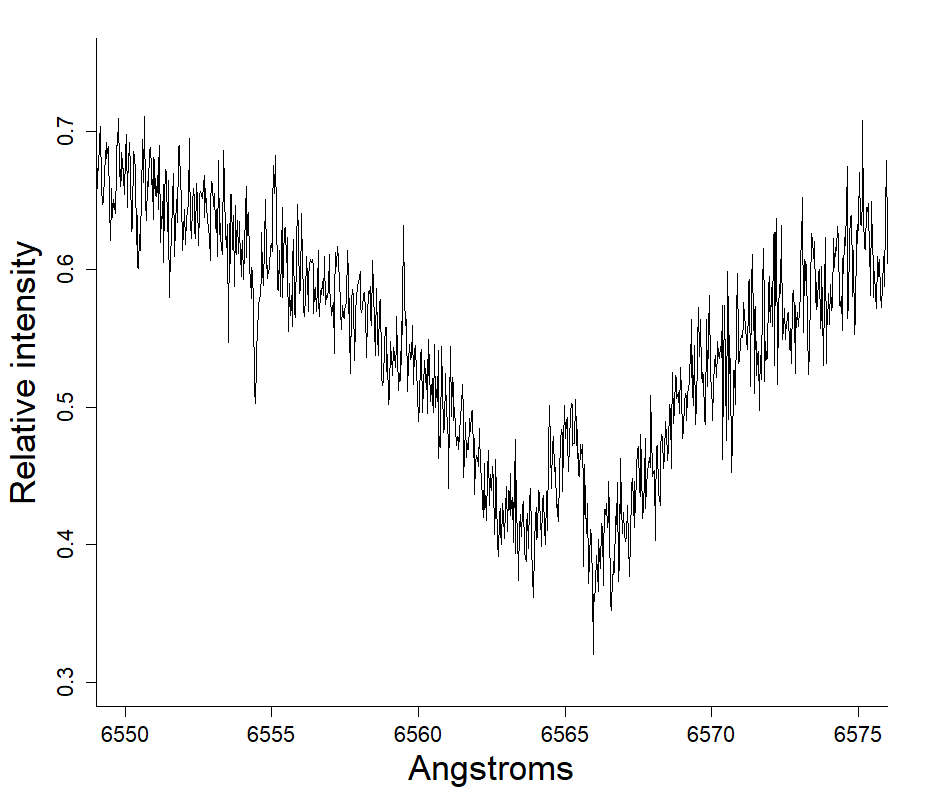}
%\vspace{0.5cm}
\caption{HERCULES order 87 showing the H$_{\alpha}$ line {of V410 Pup C} with an emission core.}
\label{StarCHalpha} 
\end{figure}

The last entry in Table~\ref{tbl-7} refers to an exposure of V410 Pup C --- the outer member of the system. We deduced from the photometry that  this star would not show any helium lines, and indeed the spectrum seems mostly featureless against a relatively noisy continuum.  {While the target is a faint (for the observational setup of this paper)}  magnitude 9.5 (Table~\ref{tab:mark_c_separate}), some comments can be made from {the observation. Low} and narrow levels of H line emission can be seen, and to a greater extent in H$_{\alpha}$. An RV of $\sim 29 \pm 2$ km s$^{-1}$ was measured for the well-defined H$_{\alpha}$ emission peak  (Figure~\ref{StarCHalpha}).

{\sc prof} estimates the equivalent width of a line by numerical  integration.  The results are given in Table~\ref{tbl-7}.  The mean value of the measured estimated widths (EWs) for the 32 inner system data-sets  in Table~\ref{tbl-7} is 0.105 $\pm$ 0.006. With a representative value of $L_{Aa}$ = 0.4 deduced from Table~\ref{tab:lc_fitting} for the relevant spectral region, {$0.26 \pm 0.1$ is} an appropriate value of the primary's $\lambda$ 6678 EW. There is a small but distinct  asymmetry in the distribution of the EWs, in that the primary-approach phases (excluding the outlier at phase 0.8 and a small number of points close to the central eclipses) show a mean EW of 0.125, while those in recession average at 0.094. The standard deviation of the complete sample is {0.036,} so the probable error of the two half-range means works out at about 0.010. The difference in the two means {is} around 3 times this value, {so} the `Struve-Sahade effect' is confirmed in these observations as a $\sim$3-sigma event.

From comparison with the calibration data of \cite{Leone_Lanzafarne_1998}, we estimate the primary's temperature to be 13000 $\pm$ 2000 K. This corresponds to a B5 spectral type: i.e., a hotter and more massive MS star than would correspond with the Michigan Spectral Survey assignation \citep{Houk_1978}. The latter  would have been affected by the significant light contributions from the other stars (Ab and B). 

\begin{figure}
\centering
\includegraphics[height=5.8cm]{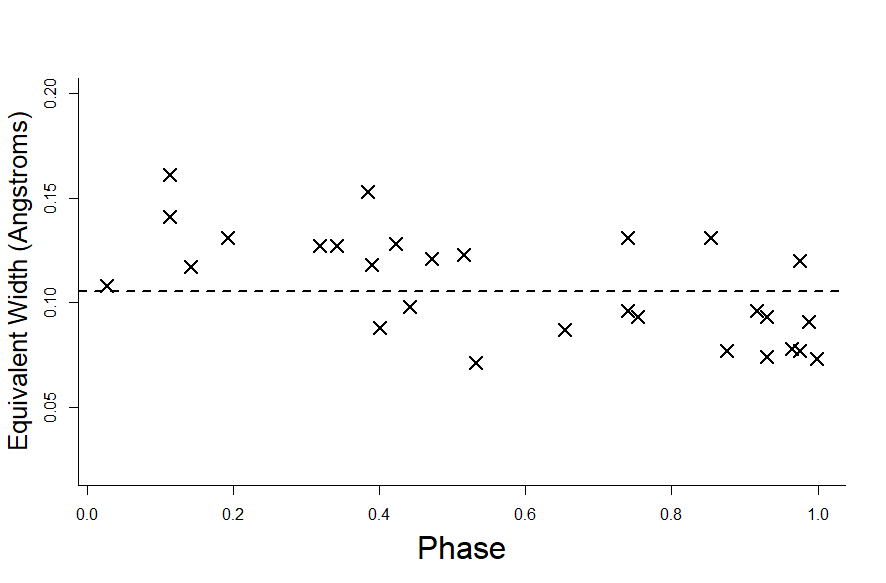}
%\vspace{0.01cm}
\caption{He I $\lambda$6678 equivalent width measures plotted against orbital phase.}
\label{fig:gaussian_fitting} 
\end{figure}

\begin{figure}
\centering
\includegraphics[height=4.9cm]{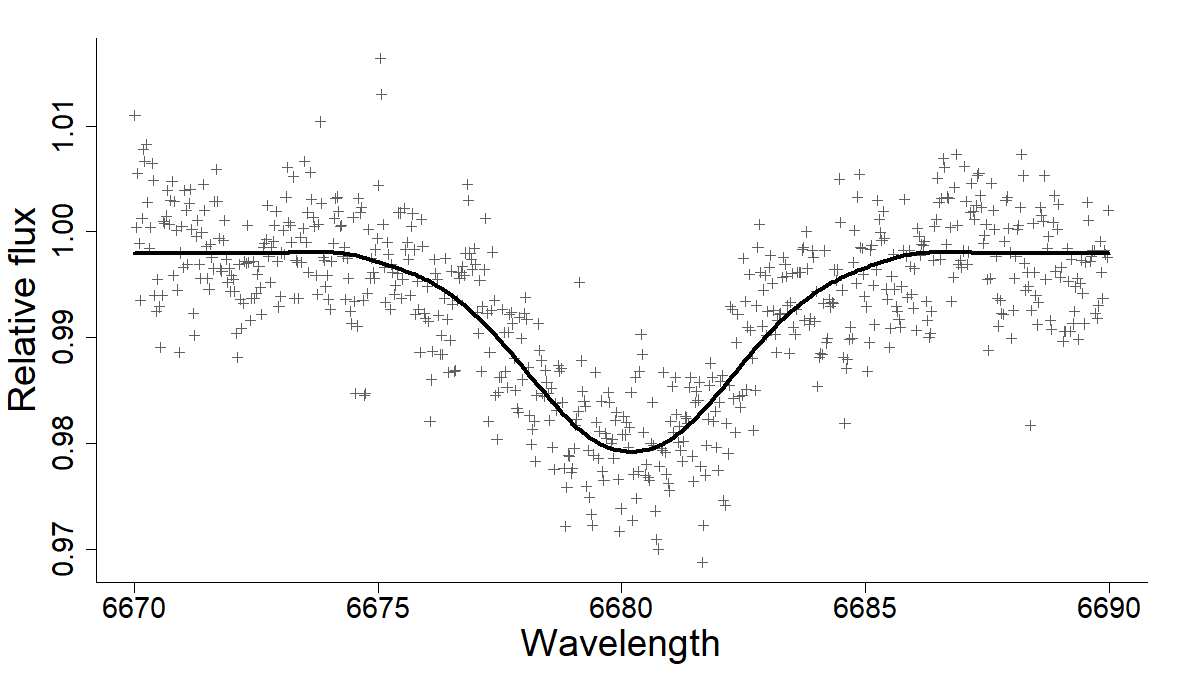}
\vspace{0.1cm}
\caption{Convolved rotation Gaussian fitting to the He I 6678 line profile.}
\label{fig:gaussian_fitting_single} 
\end{figure}

If the {stellar} rotations {(of Aa and Ab)} are synchronized to the orbit, as would be expected for this relatively very close pair \citep{Zahn_1977}, projected equatorial  speeds {would be of order} 120 km s$^{-1}$ for the primary and 85 km s$^{-1}$ for the secondary,  i.e., rotational half-widths of around 2 \AA\ at $\lambda$ 6678.  Generally, the best defined non-hydrogenic stellar absorption feature in our spectra is the He I line at $\lambda$~6679.996 ({\em in vacuo}). In Table~\ref{tab:profile_parameters} we present output parameters averaged from {\sc prof} fittings to 6 well defined examples.  EWs are shown by orbital phase in Figure~\ref{fig:gaussian_fitting}, while Figure~\ref{fig:gaussian_fitting_single} shows an example of such a fitting. %with a corrTable~\ref{tab:profile_parameters}corresponding parameter set given in 
  
In Table~\ref{tab:profile_parameters}, the reference ordinates $I_c$ and $I_d$ correspond to the mean level of the local continuum and central line depth, respectively.  The mean value of the rotation parameter $r$  yields a projected mean equatorial rotation speed of 124 $\pm$ 20 km s$^{-1}$.  This parameter is determined implicitly through the formula $x = (\lambda - \lambda_0)/r$, where $x = \pm 1$ at the equatorial limb.  In this way, $r$ scales the rotational broadening from the profile function $P(x)$ to the observed wavelengths $\lambda$,  where:  
\begin{equation}
P(x)= I_c + \frac{3 I_d}{(3 - u)} \left( (1-u) J_1(x) + \frac{u\pi}{4} J_2 (x) \right)  .
\end{equation}

Here, $J_1$ and $J_2$ are integrals forming the convolutions of the rotational  and  Gaussian broadening across the  uneclipsed stellar disk, %(for which $-1 \leq x \leq 1$), 
with dependence on an adopted  limb-darkening coefficient $u$. The integral $J_2$ has an explicit form involving the Gaussian and error functions of $x$. A simple  explicit form for $J_1$ is not known, however, so {\sc prof} evaluates this quantity by a numerical quadrature. In principle, $P(x)$ could be extended to include the broadening effects of damping, pressure and optical depth. {However the} first two are taken to be relatively small for the helium lines (compared with their rotation and turbulence broadening), and it is assumed that the features are formed at low optical depth.  Absorption scales linearly with the number of contributing atoms in the line of sight at the wavelength $\lambda$. 

Significant uncertainty remains with the fitting of  the turbulence (Gaussian) parameter $s$, {which we} associate with non-stationary effects.  \cite{Olah_1992} cited \cite{Lang_1980} for the formula: 
\begin{equation}
s = \frac{\lambda}{c}  \sqrt{ \frac{kT}{m} + \frac{v_t^2}{2}}   
\end{equation}
where symbols have their usual meanings, $v_t$ being associated with macro-turbulent motions of bulk volumes of gas. The same fittings {which provided estimates for} the rotation speed gave a value of $s \equiv 25 \pm 8$ km s$^{-1}$.
  
% For comparison purposes, a useful direct estimate of the synchronous rotation velocity
%for uniformly rotating spherical stars in circular orbits can be shown to be given by $v_{\rm sync} = %0.6 R/P$, 
%where $R$ is the stellar radius in solar units and $P$ is the orbital period in days.  This would %yield 
%%\footnote{A velocity in AU y$^{-1}$ is multiplied by 4.741 to convert it to km s$^{-1}%$.}

\begin{table}
    \caption{Main profile fitting parameters for {the} $\lambda$6678 He~I (primary) absorption  line (see text)}
    \label{tab:profile_parameters}
        \begin{center}
            \begin{tabular}{lr}
            \hline 
        %\multicolumn{2}{l}{V 410 Pup: He~I 6678 properties} \\ 
            \multicolumn{1}{c}{Parameter} & \multicolumn{1}{c}{Value} \\
            \hline 
%            \multicolumn{2}{l}{Primary} \\ 
            $I_c$                    & $0.998 \pm 0.001 $    \\ 
            $I_d$                    & $-0.021 \pm 0.005 $   \\ 
           % $\lambda_m$              & 6678.95 $\pm$0.6     \\ 
            $r$                      & $2.76 \pm 0.42$       \\ 
            $s$                      & $0.56 \pm 0.20$       \\ 
            $\chi^2/\nu$             & 1.08          \\ 
            $\Delta l$               & 0.005 \\
            \hline
            \end{tabular}
        \end{center}
%{\small The wavelengths shown are vacuum.}
\end{table}

%-------------------------------------------------------------------------------

%----------------------------------------------------------------------------

\begin{figure}
\centering
\includegraphics[height=4.9cm]{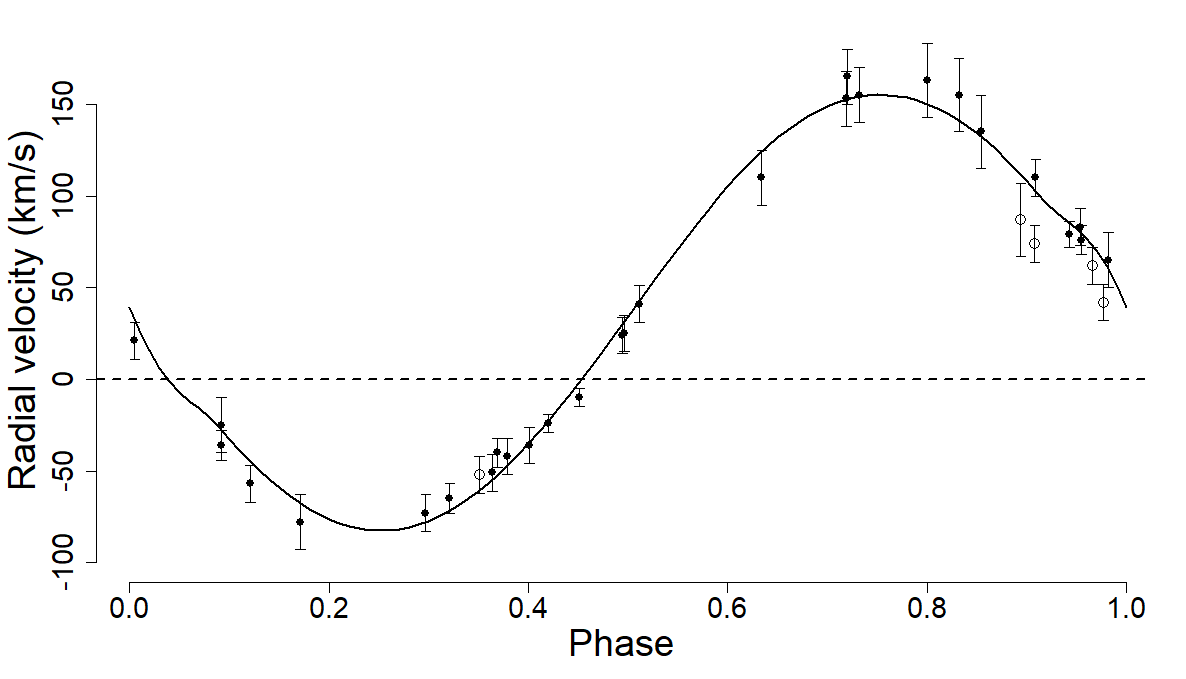}
%\vspace{0.5cm}
\caption{Radial velocity curve for the primary {(Aa)} of {central binary in} V410 Pup.  The solid curve shows the {predicted velocities based on the best-fit model.} {Its} amplitude {depends} on the mass ratio through Eqn~4.}
\label{fig:rv_fitting} 
\end{figure}

%----------------------------------------------------------------------------

\begin{table}
\begin{center}
\caption{Out-of-eclipse curve-fitting results for measured MJUO RVs {of} the close binary {(Aa, Ab)} in V410 Pup.  Symbols have their usual meanings.
\label{tab:rv_fit}} 
\begin{tabular}{lrr}
\hline 
%\multicolumn{1}{c}{}  & 
%\multicolumn{3}{c}{2011- 2014}  \\ 
%\multicolumn{1}{c}{2014} & \multicolumn{1}{c}{All} &\multicolumn{1}{c}{ }\\
\multicolumn{1}{l}{Parameter}  & \multicolumn{1}{r}{Value}  & 
\multicolumn{1}{r}{Uncertainty}\\
\hline \\
$K_{Aa} \sin i$ (km s$^{-1}$)          & 122         & 3          \\
$a_{Aa} \sin i$    R$_{\odot}$         & 2.11        & 5.4        \\
%$\Delta \phi_0$ (deg)              & 23.6         & 1.2        \\
$V_\gamma$ (km s$^{-1}$)            & 36           &  2         \\
$f(M)$ (M$_{\odot}$)                & 0.165        & 0.014      \\
$q$                                 & 0.52         & 0.01          \\
$\sigma$ (km s$^{-1}$ )             & 11           &            \\
$\nu$                               &  28          &            \\
$\chi^2/\nu$                        & 1.08         &            \\
\hline
\end{tabular}
\end{center}
\end{table} 
 
{The RV curve shown in Figure~\ref{fig:rv_fitting} is based on} the $\lambda_m$ values determined by the line-fitting process. The observed data-points were  optimally matched by the theoretical model curve (the continuous line shown in Figure~\ref{fig:rv_fitting}) {derived using the} RV-curve application of {\sc WinFitter}.  {The} corresponding parameters {are} listed in Table~\ref{tab:rv_fit}.

%----------------------------------------------------------------------------------------

\subsection{Mass function}
\label{sec:mass_function}

The well-known form of the mass-function $f(M)$ for a binary system can be written as \citep{Ludendorff_1911}:
\begin{equation}
    f(M) = C (1 - e^2)^{3/2} K_1^3 P  \,\, 
         = \,\, M_1 q^3 \sin^3 i /(1 + q)^2  \,\,\,  ,
         \label{eq:five}
\end{equation}
with symbols having their usual meanings and where $C$ is a constant $\approx 1.0361 \times 10^{-7}$ for $K_1$  in km s$^{-1}$ and $P$ in days.  For example, if V410 Pup were a semi-detached system with  $q \approx 0.5$ and the primary a Main Sequence-like star of mass $\sim$3 M$_{\odot}$, we would have $f \approx  0.138$. {$K_1$ would then be} $\approx 115$ km s$^{-1}$.  The model of an unevolved, near equal mass pair would {however} require a much higher value of $K_1 \approx 190$ km s$^{-1}$.  

This point bears strongly on our understanding of the implications of the radial velocity data. Although the value of $f(M)$ listed in Table~\ref{tab:rv_fit} is a little greater than the foregoing crude estimate, the value of $K_{Aa}$ is significantly  lower than would be required for an unevolved model with near-equal masses. We may thus countenance a semi-detached alternative resembling, from a binary evolution point of view,  V Pup \citep{Budding_2021}. The secondary fractional radii $r_{Ab}$ values in Table~\ref{tab:lc_fitting} 
%(on page~\pageref{tab:lc_fitting}) 
are, however, significantly smaller than the corresponding Roche mean radii ($\approx 0.32$).

%----------------------------------------------------------------------------------------

\subsection{Disentangling the spectra}
\label{sec:korel}

To separate the blended spectral lines of the Aa and Ab components and to uncover information on the secondary RVs, we applied  the code {\sc korel}  for Fourier disentangling \citep{hadrava95}.  Because the light of component B enters the spectrograph together with that of Aa and Ab, it is required to disentangle the  three components in a hierarchical structure. The spectroscopic observations cover only a small part of the orbit of component B around the inner binary A and the amplitude of its RV variation is small. Consequently, there is no real chance to find in the currently available  spectra either changes in RVs of the centre of mass of sub-system A and component B, or a corresponding light-time effect. We can thus treat component B as static by selecting the priors: (i) low amplitude of third star RVs, (ii) mass ratio of order unity and (iii) long orbital period.

\begin{table}
\begin{center}
\caption{Orbital parameters deduced from {\sc korel} disentanglement 
\label{Pup3par}} 
\begin{tabular}{lrr}
\hline 
\multicolumn{1}{l}{Parameter}  & \multicolumn{1}{r}{Value}  & 
\multicolumn{1}{r}{Formal error}\\
\hline 
$T_0$                               & 2456256.9190 & 0.0001    \\
$K_{Aa} \sin i$ (km s$^{-1}$)          & 133.9        & 0.4         \\
$K_{Ab} \sin i$ (km s$^{-1}$)          & 230.4        & 0.8         \\
$a \sin i$    R$_{\odot}$           & 6.30         & 0.05        \\
$q$                                 & 0.581         & 0.002     \\
%$\sigma$ (km s$^{-1}$ )             & 11           &            \\
%$\nu$                               &  28          &            \\
%$\chi^2/\nu $                       & 1.08         &            \\
\hline
\end{tabular}
\end{center}
\end{table} 

Because of the {variation in} flux due to the eclipses of the subsystem A, it is advantageous to disentangle all three components with free line-strength factors \citep{hadrava97}. In addition, the flux ratio of the components A and B can change between different exposures depending on the {telescope orientation (on the sky) and weather conditions. This is} because the angular distance between these components is not completely negligible compared to the entrance pupil of the spectrograph. This variation, however, should be small owing to the typical size of seeing, which is larger than the current separation of the components close to the periastron. Regarding the short orbital period of  the inner pair Aa and Ab, which is well established from the photometry, we can safely assume the inner orbit to be circular and solve for three orbital parameters -- (1) the time $T_0$ of primary eclipse, (2) RV amplitude $K_{Aa}$ of the primary, and (3) the mass  ratio $q=M_{Ab}/M_{Aa}$.

\begin{figure}
\centering
\includegraphics[scale=0.48]{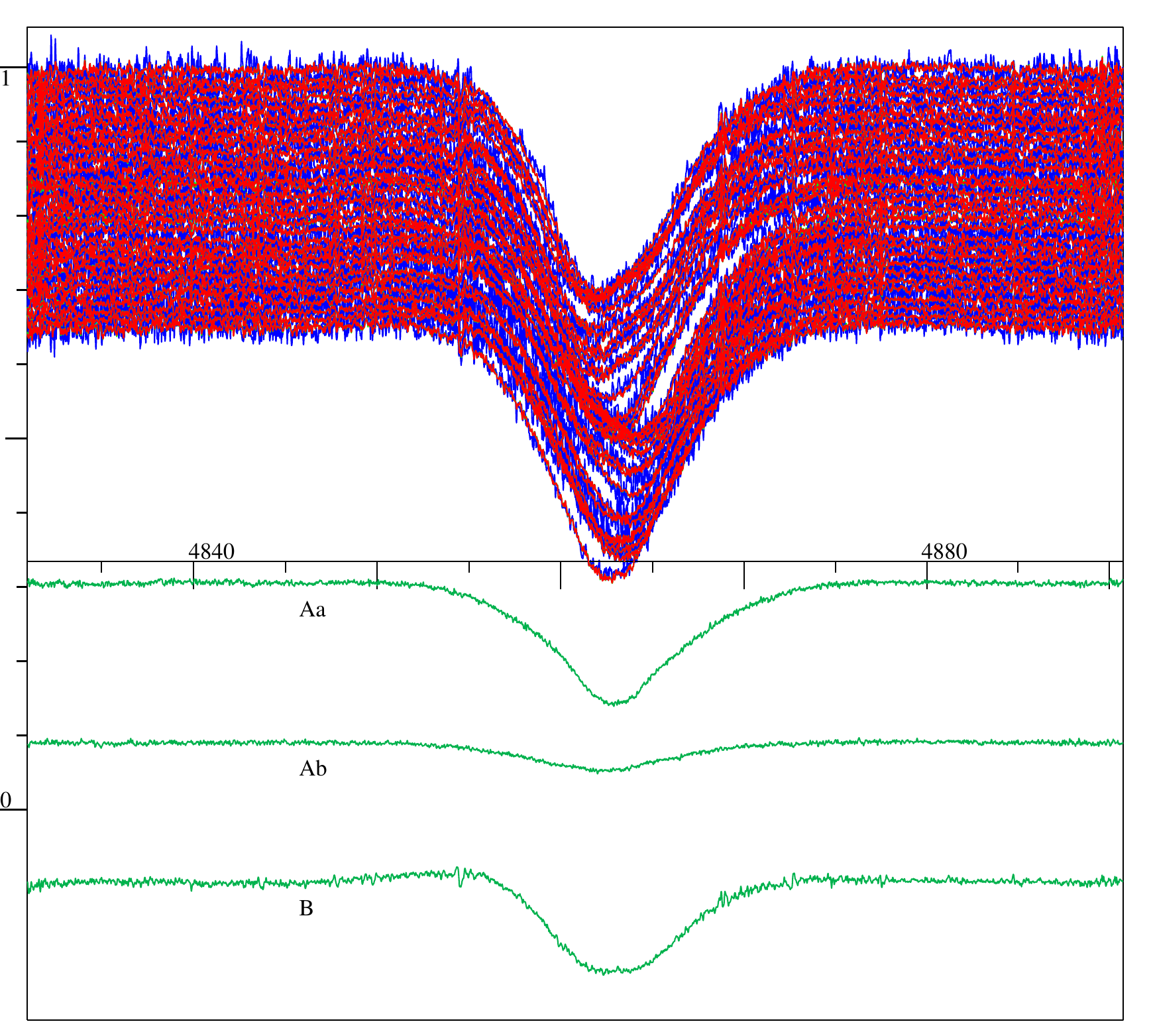}
\caption{$\rm H_\beta$ line of V410 Pup.  Intensity is on the vertical axis, and wavelength on the horizontal.  The upper panel shows the raw (blue) and fitted (red) observations. The lower panel shows the three disentangled $\rm H_\beta$ lines.}
\label{Pup3dis}
\end{figure}

\begin{table}
\caption{{\sc korel} RV determinations of the components of V410 Pup. BJD is the  Barycentric Julian Date of the observations. The stellar component is indicated by the {subscript, Aa for the primary and Ab} for the secondary. The $(O-C)$ columns refer to differences between the observed versus the theoretical model, as shown in Figure~\protect{\ref{rvs}}.}
\label{tableA1}
\begin{tabular}{crrrr}
                    &                   &                   &               &            \\
\hline
BJD			        & $RV_{Aa}$		    & $(O-C)_{Aa}$		& $RV_{Ab}$	    & $(O-C)_{Ab}$		\\
($-2400000$)   		& (km/s)			& (km/s)			& (km/s)		& (km/s)			\\
\hline
55879.9288	        &	131.77	        &	0.08	        &	$-226.58$	&   	$-0.10$	\\
55880.1346	        &	36.96	        &	0.09	        &	$-63.45$	&	$-0.08$	\\
55880.8750	        &	127.15	        &	$-0.07$	        &	$-218.56$	&	0.20	\\
55880.9035	        &	116.05	        &	0.00	        &	$-199.59$	&	$-0.02$	\\
55883.0634	        &	$-127.75$	    &	0.02	        &	219.70	    &	$-0.09$	\\
55883.0845	        &	$-120.20$	    &	0.07	        &	207.45	    &	0.58	\\
55883.1259	        &	$-97.68$	    &	0.08	        &	168.70	    &	0.54	\\
56257.8766	        &	$-74.10$	    &	0.02	        &	132.57	    &	5.08	\\
56257.9458	        &	$-118.35$	    &	$-0.06$	        &	203.23	    &	$-0.25$	\\
56259.0672	        &	$-39.47$	    &	0.12	        &	68.21	    &	0.09	\\
56259.1049	        &	$-3.81$	        &	0.17	        &	6.62	    &	$-0.24$	\\
56667.0650	        &	$-93.21$	    &	0.01	        &	160.34	    &	$-0.03$	\\
56669.9055	        &	$-100.83$	    &	$-0.13$	        &	173.23	    &	$-0.01$	\\
56669.9385	        &	$-77.66$	    &	$-0.47$	        &	132.75	    &	$-0.05$	\\
56670.8979	        &	$-2.26$	        &	0.05	        &	5.00	    &	0.99	\\
56671.8944	        &	100.17	        &	$-0.09$	        &	$-172.41$	&	0.00	\\
56673.9285	        &	36.45	        &	0.06	        &	$-62.54$	&	0.01	\\
57351.1053	        &	104.93	        &	$-0.75$	        &	$-181.68$	&	0.05	\\
57352.0590	        &	46.34	        &	0.03	        &	$-79.65$	&	$-0.03$	\\
57355.0698	        &	$-91.85$	    &	$-0.31$	        &	157.28	    &	$-0.20$	\\
57356.0614	        &	9.84	        &	$-0.25$	        &	$-17.25$	&	0.07	\\
57357.1201	        &	131.43	        &	$-0.20$	        &	$-227.05$	&	$-0.68$	\\
57358.0074	        &	132.75	        &	$-0.44$	        &	$-228.77$	&	0.28	\\
57359.9993	        &	$-5.82$	        &	$-0.34$	        &	9.60	    &	0.14	\\
57360.9513	        &	$-74.11$	    &	$-0.17$	        &	127.20	    &	$-0.01$	\\
58887.9102	        &	82.18	        &	0.18	        &	$-141.83$	&	$-0.85$	\\
58887.9225	        &	72.42	        &	0.07	        &	$-124.46$	&	$-0.08$	\\
58887.9729	        &	27.79	        &	$-0.03$	        &	$-47.91$	&	$-0.12$	\\
58887.9831	        &	18.19	        &	0.03	        &	$-32.71$	&	$-1.52$	\\
58890.9390	        & $-107.14$         &	0.15	        &	184.64	    &	0.05	\\
\hline
\end{tabular}
%\end{center}
\end{table}

For the disentangling we used 30 exposures in the spectral region 4831 -- 4891\,{\AA}  around the $\rm H_\beta$ line. This region {was} sampled in 4096 bins with a step {size} corresponding to RV $\sim$0.9 km/s. The separated line-profiles of all three components are displayed in green lines in  the lower part of Fig.~\ref{Pup3dis}. The observed spectra are shown by blue lines in the upper  part of the figure. They are overplotted in red by their reconstruction from the superposition of the separated spectra, {which are} Doppler-shifted and amplified {appropriately for} each exposure. The residual noise  is $\sim$0.01 of the normalized continuum level.  Component B contains several narrow features which seem to be of an instrumental origin  and hence are imprinted to the component with {small} Doppler shifts. The line-strength factors of components Aa and B are  anticorrelated with a correlation coefficient $-0.84$, corresponding to the varying flux ratio of these components. This trend is hardly noticeable for the component Ab,  given the relatively large uncertainties of its line-strength factors, due to the inherent shallowness of its line-profile.

The values of disentangled parameters with their Bayesian errors \citep{hadrava16} are given in Table~\ref{Pup3par}. These errors indicate how the fit for each model parameter  can be influenced by photon noise. They are thus termed formal errors.  Usually they are smaller for high S/N spectra than differences between parameter values obtained from different lines. Low formal errors arise because the model only takes into account the Keplerian Doppler shift and the line-profile changes in the line-strength factors. It neglects other effects, such as the tidal distortion, reflection, the influence of circumstellar matter or other such complications.  In this way, line profile constructions need not follow precisely the motion of the stellar centres of mass.

\begin{figure}
\centering
\includegraphics[scale=0.39]{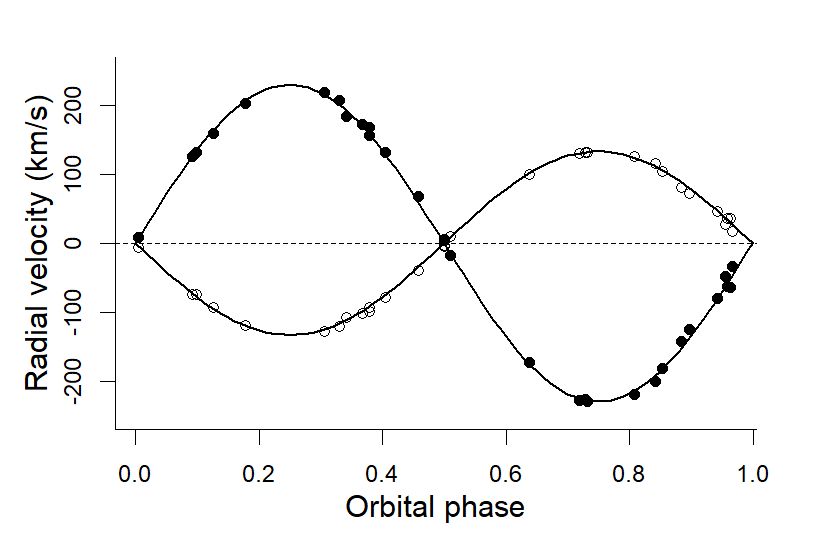}
\caption{{\sc korel} radial velocities from $\rm H_{\beta}$ lines of V410 Pup and WD model fitting.}
\label{rvs}
\end{figure}

The radial velocities obtained in {\sc korel} by fitting each exposure independently as a superposition of the disentangled profiles have {the following root mean square} errors (i.e., deviations from the RVs given by the disentangled orbital parameters) {of} 0.2 km/s for the primary (Aa), 1.0 km/s for the secondary (Ab), and 0.4 km/s for the component B. The relatively large error for the component Ab is due to its shallow profile. The deviations are particularly high in certain exposures, where this component has a very small s-factor. The {\sc Korel} RV determinations of the components of V410 Pup A are given in Table \ref{tableA1} and are shown in Fig. \ref{rvs}. 

The disentangled H$_\beta$ profile of component B is of evidently different shape to the other two: its core region is widened out, suggestive of a high rotation, while the wings are curtailed, as if filled in by diffuse emission.  This is more noticeable on the blue side of the line, giving the third profile a distinct asymmetry compared to the other two.  This has some resemblance, at a weaker level, to  the wide orbit component in the triple star $\nu$ Gem, shown by \cite{Jaschek_1987} in their Figure 9.9d as an example of a Be star effect.  A recent  discussion of this system was given by \cite{Klement_2021}, where we note,
in this connection, Figure B1 of that paper.  Another example is the triple Be system 66 Oph \citep{Marr_2021} whose wide component similarly demonstrates fast rotation.

Taking these properties in conjunction with the increase in luminosity to the red, we propose that star B is surrounded by a  weak accretion structure.
 
%{\color{red}There is an interesting analogy
%(as I mentioned recently together with 66 Oph), that the fast %rotating
%component is the outer one in the hierarchical triple. So, perhaps, %it
%would be interesting to mention.}

%------------------------------------------------------------------------------------------------

\section{Astrometry}
\label{sec:astrometry}

{Investigation of astrometric positions is highly} relevant for young massive stars in the nearer star-forming regions, given the likely presence of components at angular separations in the order of tens of mas. Outcomes include orbit sizes, nodal angles and inclinations, as well as, from Kepler's laws, masses of the stars.

The {\em Washington Double Star Catalogue} \citep{Mason_2010} lists 19 position angle ($\theta$) and separation ($\rho$)  measures for the V410 Pup AB system (AB = I 1070, \citealt{Innes_1899}), as well as 14 measures of the AB-C system (HJ 4032, \citealt{Herschel_1836}).  {The source also appears in the cross-calibration of Hipparcos and Gaia EDR3 by \cite{Brandt_2021}.}

These measures can be referred to standard formulae for the co-ordinates of the relative orbit \citep{Ribas_2002}: 
\begin{equation}
X = \frac{a(1 - e^2)}{1 + e \cos \nu}\bigg\{\cos(\nu + \omega)\sin\Omega +
\sin(\nu + \omega)\cos\Omega\cos i\bigg\} ,
\end{equation}
and
\begin{equation}
Y = \frac{a(1 - e^2)}{1 + e \cos \nu}\bigg\{\cos(\nu + \omega)\cos\Omega -
\sin(\nu + \omega)\sin\Omega\cos i\bigg\} .
\end{equation}
{The} equations are related to the differential positional measurements $\rho$ and $\theta$, since 
$ \rho = \sqrt{X^2 + Y^2}$ and  $\tan \theta = X/Y $;
so that the measures are directly related to the 5 orbital parameters $a$, $e$,  $\omega$, $i$, and $\Omega$. To fix the configuration at a given time, we also need the period $P$ {and} the epoch of periastron passage $T_0$, a provisional value for $T_0$ for {component B} can be estimated from the shape of the apparent orbit, which appears to {currently be} close to periastron.

{Optimization is a practical method to estimate the parameter values} in which, {by} using an appropriate search algorithm, {an orbital ellipse based on the parameter estimates} is  progressively matched to the data so as to minimize residuals (see Chapter 5 of \citealt{budding_demircan} and \citealt{Rhodes_Budding}).  The procedure  involves the seven aforementioned parameters.
%  as well as small translational corrections
% $\Delta x_0$, $\Delta y_0$, in the position to be assigned to the origin.
Formally, we can write for the solution of the inverse problem:
\begin{equation}
{\bf a}_{\rm opt} = [\chi^2]^{-1}{\rm Min}[\chi^2 ( {\bf a} ) ]  \,  ,
\end{equation}
where ${\bf a}_{\rm opt}$ is the vector of best estimates of each parameter in the adjustable set $\{ a_1, a_2, a_3 ... a_m \}$.  
 %The observed values of 
 %the variable, which could be $\rho$ or $\theta$, 
%are matched by the values of the fitting function. 
The quantity $\chi^2$ depends on the squared differences of observed and calculated quantities, and the optimal estimate  for each  $a_j$ is taken to occur when $\chi^2$ is minimized. $[\chi^2]^{-1}$ expresses functionally  the  inverse nature of the operation.  

\begin{figure}  
\centerline{\includegraphics[height=7.5cm]{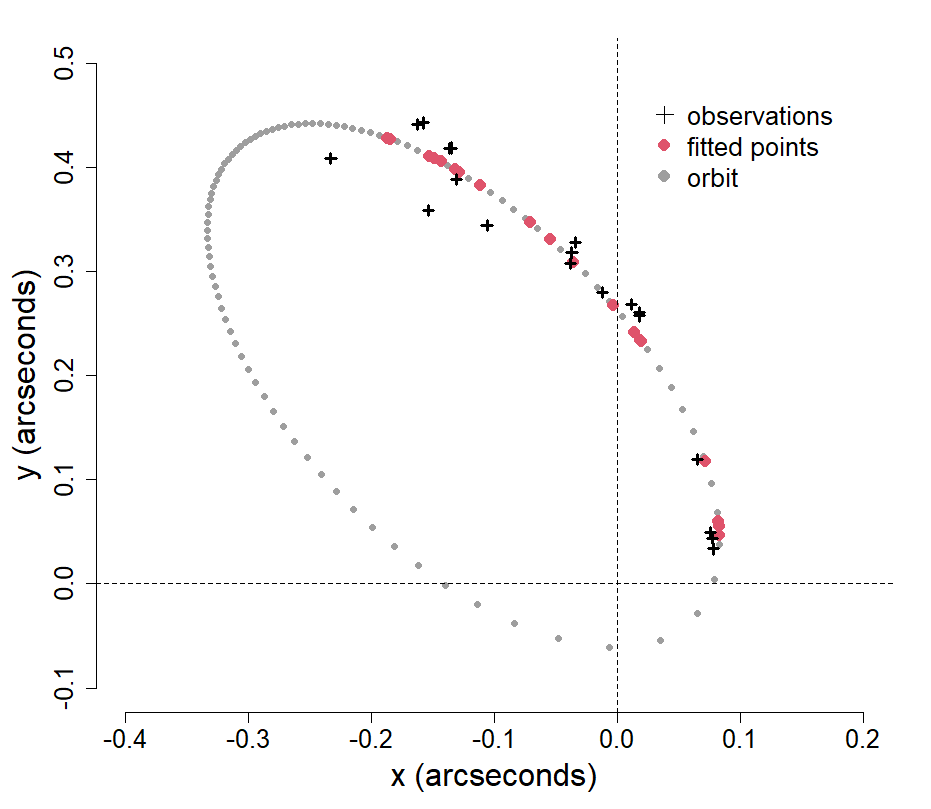}}
%[trim=0cm 14cm 0cm 4cm, clip=true, height=5cm, width=10cm]{images/v4109ast24.pdf}}
\caption{Fitting to the astrometric orbit of I 1070 (components A and B). The open triangles represent the observations of separation and position angle. Corresponding orbital positions are indicated by the sequence of full circles.
\label{fig:I1070_orbit}}
\end{figure}

The adopted result for V410 Puppis B is shown in Figure~\ref{fig:I1070_orbit}, and the corresponding parameters listed in Table~\ref{tab:I1070}, together with indicative error estimates. With only a third of the  orbit completed and, understandably low precision of the early data, the determinacy of the model is not high. The results of Table~\ref{tab:I1070} may therefore be regarded as provisional. This concerns the correlation between parameters ({important for the purpose of finding masses)} inclination $i$ and eccentricity $e$.  A higher $i$ to a wide orbit, for a given apparent eccentricity, entails a lower true value of $e$; so the greater travel implied in the fixed period requires higher total mass.  The derived masses thus have relatively large uncertainties, but a moderately low $i$  produces masses in keeping with those of the system components found in the preceding sections.

%It is interesting that a good-fitting inclination for the A-B system appears close to that of the close binary. 
%Within the errors of measurement we could say that the V831 Cen ab and
%See 170 systems are consistent with coplanar orbits.

%--------------------------------------------------------------------------------
% Cropping: “trim=1cm 2cm 3cm 4cm” trims (crops) from left, bottom, right and top by 1, 2, 3 and 4cm respectively. It must be accompanied by “clip=true”.

%--------------------------------------------------------------------------------
 
\begin{table}  
\caption{Astrometric elements for I 1070 (Components AB of V410 Pup)}
\label{tab:I1070}
\begin{center}
\begin{tabular}{lrr}
\hline
\multicolumn{1}{l}{Parameter} & 
                        \multicolumn{1}{r}{Value} & 
                                              \multicolumn{1}{r}{Uncertainty}\\
\hline
$P$ (y)                 & 379                 & 35      \\ 
$a$ (mas)               & 297                 & 21      \\ 
$i$ (deg)               & 24                  & 14      \\ 
$\Omega$ (deg)          & 120                 & 15      \\ 
$T_0$ (y)               & 1650                & 35      \\ 
$e$                     & 0.80                & 0.09    \\ 
$\omega$ (deg)          & 26                  & 16      \\ 
$M_{\color{blue}Aa}+M_{\color{blue}Ab}+M_{\color{blue}B}$ ($\odot$) & 8                   & 3       \\ 
\hline
Fitting stat.           & $\chi^2/\nu$ = 1.01 & $\Delta s$ = 35 (mas)   \\ 
\hline
\end{tabular}%
\end{center}
\end{table}
%--------------------------------------------------------------------------------

The semi-major axis, at the Gaia distance of 346 pc, produces a mean physical separation of $\sim$110 AU for V410 Pup AB. Kepler's law then yields  $\sim$8 $ M_{\odot}$ for the total mass of the system. The photometric + spectroscopic solutions suggested masses of 3.15 and 1.83 $M_{\odot}$ for the close binary (A), so the total mass of the A-B system is in agreement with this if the third star is a $\sim$ 3.02 $M_{\odot}$ object. A larger distance to the system (see below) would increase the mass of V410 Pup B.

The radial velocity of star B relative to A, according to the orbit model of Figure~\ref{rvs}, corresponds at the observed phase of about 0.9 to about 12.0 km s$^{-1}$ in approach (using formula 9.8 in \citealt{budding_demircan}).  The spectrographic mean RV of $36 \pm 2$ km s$^{-1}$ for the close binary (Table~\ref{tab:rv_fit}) is then an acceptable $\sim$6 km s$^{-1}$ greater than the mean velocity (+30 km s$^{-1}$) of the AB system.

In addition to observations of V410 Pup AB, \cite{Mason_2010} also provided observations for HJ 4032 ({star C in} the AB-C system). These trace out an almost linear short trend when plotted; implying that this limited coverage of {C's} orbit could not be used to establish a full set of orbital parameters. The data {cover} only about 180 years of a period that could well be in the order of five thousand years.

\section{Absolute parameters}
\label{sec:absolute_parameters}

The well-known `eclipse method' combines photometric and spectroscopic findings, together with the use of Kepler’s third law, or other available  relationships, to derive basic physical parameters of the component stars. 
% We deduced a mass ratio of $q \approx 0.5$ from Section~\ref{sec:mass_function}, %and that can be put more precisely if we adopt a Main Sequence-like mass for the 
% primary star of 3.1 M$_{\odot}$ and solve equation~\ref{eq:five} 
%(on page~\pageref{eq:five}) 
%iteratively. The derived value of $q$ ($\approx 0.53$)  is not very sensitive  to the value of $M_1$, but increases somewhat with a lower $M_1$. The projected  average distance between the primary and the system centre of mass $a_1 \sin i$ is found from multiplying $K_1$ by the period  in seconds and dividing by $2\pi$ for  a circular orbit. The semi-major axis of the relative orbit $a$  follows  by multiplying this number by $(1+ q)/(q \sin i) $ and the result is given in Table~\ref{tbl-10}, where solar units are used, unless otherwise indicated.
{Taking} the orbital inclination from the WD+MC solution in Table~\ref{tab:lc_fitting}, {the} average distance between components from the RV solution in Fig.~\ref{rvs}, and {the calculated} period {all in an application of} Kepler’s third law, {then} the combined mass of the close components (Aab) is found to be $4.97\pm0.18$ M$_{\odot}$. The masses of the individual components are then calculated from the spectral mass-ratio ($q=0.58$).

%-------------------------------------------------------------------------------
\begin{table}
\caption{Absolute parameters of V410 Pup (solar units).
\label{tbl-10}}
\begin{center}
\begin{tabular}{lrr}
\hline
\multicolumn{1}{l}{Parameter} & \multicolumn{1}{r}{Value} & 
\multicolumn{1}{r}{Uncertainty} \\ 
%\multicolumn{1}{c}{solar units} & \multicolumn{1}{c}{ } & 
%\multicolumn{1}{c}{ } \\
\hline
$M_{Aa}$                   & 3.15              &   0.10        \\ 
$M_{Ab}$                   & 1.83              &   0.08        \\
$M_{B}$                    & 3.02              &   0.20        \\
$a_{Aab}$                  & 6.57              &   0.04        \\
% $P_{AB} y$                            & 380               &  30           \\
$R_{Aa}$                   &  2.12             &   0.10        \\ 
$R_{Ab}$                   &  1.52             &   0.08        \\
$R_{B}$                    & 3.62              &   0.20        \\
$T_{Aa}$ (K)               & 12500             &   1000        \\ 
$T_{Ab}$ (K)               & 9070              &   800         \\
$T_{B}$ (K)                & 7500              &   1000        \\ 
$\log L_{Aa}$              &  2.00             &   0.17        \\
$\log L_{Ab}$              &  1.15             &	 0.15       \\
$\log L_{B}$               &  1.57             &   0.18        \\
M$_{\rm{bol}_{\it Aa}}$    &  $-0.23$          &   0.43        \\
M$_{\rm bol \it _{Ab}}$    &  1.88             &   0.48        \\ 
M$_{\rm bol \it _{B}}$     & 0.82              &   0.45        \\ 
M$_{V_{Aa}}$               &  0.63             &   0.43        \\
M$_{V_{Ab}}$               &   2.00            &   0.48        \\ 
M$_{V_{B}}$                &  0.82             &   0.20        \\
$\log{ g_{Aa}}$            &  4.28             &   0.05        \\
$\log{ g_{Ab}}$             &  4.34             &   0.06        \\ 
$\log{ g_{B}}$             & 3.96              &   0.40        \\ 
\hline
\end{tabular}
\end{center}
\end{table}
%-------------------------------------------------------------------------------

%From Kepler’s third law, the combined mass of the close components (Aab) is found to be 4.7 M$_{\odot}$, which accords with the masses listed in Table~\ref{tbl-10} derived from the  adopted MS character of the primary and the derived mass ratio. 
Using the known separation $a$ of the close binary stars,
the fractional radii of components ($r_{Aa}$, $r_{Ab}$) obtained from WD+MC solution in Table~\ref{tab:lc_fitting} lead to the absolute radii ($R_{Aa}$, $R_{Ab}$). Surface gravities ($g_{Aa}$, $g_{Ab}$) are then directly derived. Determination of the bolometric magnitudes ($M_{\rm bol}$) and luminosities ($L$) of the component stars requires the effective temperatures that were determined from the colours in Table~\ref{tab:mark_c_separate}. %(page~\pageref{tab:mark_c_separate}).  
In these calculations, the solar calibration values are adopted as: effective temperature $T_e$ = 5772 K, $M_{\rm bol}$ = 4.755  mag,  $\rm BC = -0.107$  mag and $g_{\odot} = 27423  \: {\rm cm/s^2}$, following the listings of \cite{Pecaut_Mamajek_2013}, with the bolometric corrections (BC) for the components taken from the tabulation of \cite{Flower_1996} according to the assigned effective temperatures.  Our derived absolute parameters for the V410 Pup system are listed, with their uncertainties, in Table~\ref{tbl-10} (on page~\pageref{tbl-10}). 

The anomalous value of $T_{B}$ is clearly at variance with its mass, if star B had a MS character. This finding may be associated with the putative  disk, given the  {\sc korel} results shown in Fig~\ref{Pup3dis}, and the colours of Table~\ref{tbl-4}. 
%(page~\pageref{tbl-4}).

\begin{figure}
\centering
\centerline{\includegraphics[height=6.0cm]{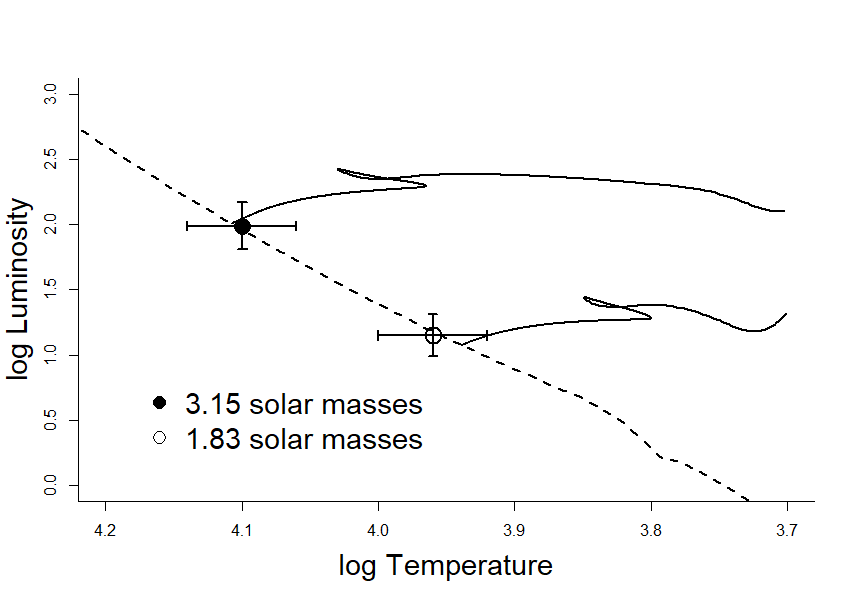}}
\caption{Location of the components {(Aa and Ab)} of V410 Pup in {the} HR diagram. The Padova evolutionary tracks \protect\citep{Bressan_2012} for 3.15 M$_{\odot}$ and 1.83 M$_{\odot}$ main sequence stars are plotted for $Z = 0.017$.  The Padova isochrone line of 25 Myr is indicated by the dashed {line} in the diagram.}
\label{fig:hr_diagram}
\end{figure}

%The HR Diagram in FIg 15 shows
{Fig~\ref{fig:hr_diagram} plots evolutionary tracks as derived from Padova modelling (\citealt{Bressan_2012}, \citealt{Pastorelli_2020}), set with initial metallicity $Z = 0.017$.  
%The diagram compares the relatively accurate stellar radii  (R$_i$) with those of the models. The theoretical early variations of radius with $\log$ age are plotted  for primary  and secondary  stars with masses 3.1 and 1.6 M$_{\odot}$.  These are compared with our derived values  (R$_i$) listed in Table~\ref{tbl-10}.   The uncertainty level of R$_1$ is shown by the dotted line at at R = 1.95 (upper bound only displayed).   The secondary's derived radius corresponds to the long-dashed line at R$_2$ = 1.87 R$_{\odot}$, with its uncertainty indicated by the dotted line at 1.81 R$_{\odot}$ (lower bound only).  Having such possible errors, the primary is consistent with any age in the range 4 to 12 My. The  secondary, however, within a short mass range near to 1.6 M$_{\odot}$, shows relatively large swings in size, associated with the onset of the C-N-O nuclear burning in its core. The star should thus be confined to the relatively narrow age range 4 to 7 My (leftward of the vertical black line) if it is to agree with the models.
This HR diagram shows evolutionary tracks for the Aa and Ab components calculated by interpolating between models taken from \protect\cite{Bressan_2012} together with surface temperature and luminosity parameters from the WD + MC findings in Table~\ref{tab:lc_fitting} %(p.~\pageref{tab:lc_fitting}), 
and the {\sc korel} results in Table~\ref{Pup3par} (p.~\pageref{Pup3par}). The WD + MC fitting gives a lower ratio of radii than the optimal {\sc WinFitter} fit to the Sector 35 TESS data, while the {\sc korel} analysis indicates a somewhat higher mass ratio than was apparent from the mass function and $q$-search methods. These two results together allow marginal conformity of Aa and Ab to near-ZAMS positions. Star Ab appears significantly hotter and more luminous for its derived mass, suggesting a still condensing structure. }

%----------------------------------------------------------------------------------------------

\section{Conclusions}
\label{sec:conclusion}

In this paper we have combined new spectrometric and Earth-based BVR observations {together} with satellite photometry {(particularly TESS data)} and modern data-processing techniques  to produce credible absolute parameters of the close binary V410 Pup Aa-Ab.  We included astrometric data analysis that reveals properties of the third major component V410 Pup B.  

We have discovered a pulsation behaviour consistent with a slowly pulsating B-type (SPB) star. {We associate this} with the Aa (spectral-type B5) component.  The magnitude and colour found for star B correspond to an object distinctly {more} cool and less massive than Aa, but we argue that B is affected by an accretion structure evidenced by its  line profile and redward luminosity excess. We deduce from the relative brightness and colour of V410 Pup C that it is an A2e type Main Sequence dwarf with a mass around 2 M$_{\odot}$.

The system has sky-location, distance, and proper motion values ($\mu_{\alpha} \cos \delta = -5.716 \pm 0.353$ and  $\mu_{\delta}$ =  $9.134 \pm 0.379$ mas y$^{-1}$; \citealt{gaia_edr3}) consistent with membership of the  Vela OB2 association, particularly with the `outer ring' of  the condensation around $\gamma$ Vel \citep{Jeffries_2009}.

That the system may prove a useful laboratory for the early stages of stellar evolution can be gathered from comparisons with recent stellar structural models.  \protect\cite{Jeffries_2017} found the mean age of the Vela OB2 association to be approximately 7 Myr from optical photometry, though their measured Li-depletion for the surrounding $\gamma$ Vel cluster produced older ages in the range 18–21 Myr.  Although the age estimated from locating Aa and Ab in the H-R plane is older than that derived from their radial expansion, the H-R age of $25 \pm 10$ Myr falls within the margins of error of the Li-depletion age range found by \protect\cite{Jeffries_2017}. 

%In this way, the detailed modelling from the Padova school solves the scenario problem raised in section~\ref{sec:photometry} as to whether a `Main Sequence' or `Algol' configuration should be  relevant for the explanation of the parameters characterizing the eclipsing binary.  It turns out that, while the preferred status is that of pre-interactive binary evolution, a simple Main Sequence generalization for the relative luminosities is inappropriate as there is a relatively large increase in radius between 5 to 9 My for a 1.6 M$_{\odot}$ star. 

The {current physical status} of stars B and C can also be fitted into this stellar youth scenario.  The third spectrum produced in the {\sc korel} disentangling is understood to result from  emission filling in the H$_{\beta}$ line above a rapidly rotating condensation.  The emission arises  from circumstellar material, plausibly in the form of a protoplanetary disk.  Component B is thus deduced to be a  relatively young T Tau-like star,  or perhaps an Ae type Herbig object (\citealt{Appenzeller_1989}; \citealt{Perez_1997}). Star C similarly shows hydrogen line emission, but with a lower degree of filling of the underlying absorption line. 

%However, if WD+MC solution given in Table \ref{tab:lc_fitting} and the KOREL RV solution given in Table \ref{Pup3par} is considered, the absolute parameters of the system change slightly: $M_1 = 3.15$, $M_2 = 1.83$ M$_{\odot}$, $R_1 = 2.10$ and $R_2 = 1.52$ R$_{\odot}$. Accordingly, the diagram showing the locations of the components in 

The combined evidence on V 410 Pup thus favours a young age, most probably in the range 7 - 25 Myr, though circumstantial factors including the SPB variation and accretion structures associated with anomalous colours and line profiles compromise  precise parametrization. Further and more detailed data-collection and modelling for the whole system  is called for to elucidate  this intriguing multiple star.  

%--------------------------------------------------------------------------------
% Cropping: “trim=1cm 2cm 3cm 4cm” trims (crops) from left, bottom, right and top by 1, 2, 3 and 4cm respectively. It must be accompanied by “clip=true”.
%\begin{figure}  
%\centerline{\includegraphics[height=6cm]{images/evrad3a.png}}
%[trim=0cm 14cm f0cm 4cm, clip=true, height=5cm, width=10cm]{images/v4109ast24.pdf}}
% Must protect the cite command in the caption as it is fragile.
%\caption{
%The diagram shows the early evolution paths of radius versus $\log$ age  for the primary R$_1$ and  secondary R$_2$ of 3.1 and 1.6 M$_{\odot}$ stars according to the modelling of \protect\cite{Bressan_2012}. These are compared with the values derived from observations listed in Table~\ref{tbl-10} (the two horizontal lines).  The long dashed line at R$_2$ = 1.87 corresponds to the secondary.   Within likely errors, the primary is consistent with any age in the range 4 to 12 My. Due to the relatively large swings in size of a 1.6 M$_{\odot}$ star in its early stages, the secondary should be confined to the age range 4 to 7 My (to the left of the vertical black line). 
%\label{fig:evrad}  
%}
%\end{figure}

%--------------------------------------------------------------------------------

%--------------------------------------------------------------------------------

\section*{Acknowledgments}
 
Generous allocations of time on the 1m McLennan Telescope and HERCULES spectrograph at  the Mt John University Observatory in support of the Southern Binaries Programme  have been made available through its TAC and supported by its  Director, Dr.\ K.\ Pollard  and previous Director, Prof.\ J.B.\ Hearnshaw.   Useful help at the telescope were provided by the MJUO management  (N.\ Frost and previously A.\ Gilmore \& P.\ Kilmartin). Considerable assistance with the use and development of the {\sc hrsp} software was given by its author Dr.\ J.\ Skuljan. We thank Prof.\ Tim Bedding (University of Sydney) for his suggestions on the interpretation of the pulsation data and advice. Encouragement and support for this programme has been shown by the School of Chemical~\& Physical Sciences of the Victoria University of Wellington, as well as the Royal Astronomical  Society of New Zealand and its Variable Stars South section  (http://www.variablestarssouth.org).  The Astronomical Institute acknowledges support from the Czech Republic Academy of Sciences (RVO 67985815).  It is a pleasure to express our appreciation of the high-quality and ready availability,  via the Mikulski Archive for Space Telescopes (MAST), of data collected by the TESS mission.  Funding for the TESS mission is provided by the NASA Explorer Program.  This research has made use of the SIMBAD data base, operated at CDS, Strasbourg, France, and of NASA’s Astrophysics Data System Bibliographic Services.  We thank the University of Queensland for collaboration software, {and also the anonymous referee for guidance which improved the final paper.}

\section*{Data Availability}
The majority of data included in this article are available as listed in the paper or from the online supplementary material it refers to. The first exception is the BVR photometric data which will be shared on reasonable request to the corresponding author. The second exception is the spectroscopic imagery. We are currently working to store the unprocessed imagery associated with the Southern Binaries Project (of which V410 Puppis is part), and will be pleased to share details with requestors when that work is complete. The HIPPARCOS and TESS data are available online from the MAST and ESA\footnote{http://www.rssd.esa.int/Hipparcos/catalog.html} repositories. 

%-------------------------------------------------------------------------
    
{}

%%%%%%%%%%%%%%%%% APPENDICES %%%%%%%%%%%%%%%%%%%%%
% Do not delete the material below, it took ages to work out to get this to work.  We might
% need it again in the future.
%\clearpage
%\onecolumn
%\appendix
%\section{Table of radial velocity measurements}
%\label{secA}
%\setcounter{table}{0}
%\renewcommand{\thetable}{A\arabic{table}}
%\FloatBarrier

% Don't change these lines
\bsp	% typesetting comment
\label{lastpage}
\end{document}